\documentclass[a4paper,11pt]{article}
\pdfoutput=1 

\usepackage{jheppub} 
\usepackage{physics}
\usepackage{amsmath}
\usepackage{bbm}
\usepackage{xcolor}
\usepackage{mathtools}

\usepackage{tensor} 

\newcommand\fft[2]{\frac{#1}{#2}}

\newcommand\nn{\nonumber}

\usepackage[T1]{fontenc} 

\preprint{LCTP-19-14}

\title{The Scales of Black Holes with nAdS$_2$ Geometry}

\author{Junho Hong,}
\author{Finn Larsen,}
\author{James T. Liu,}

\affiliation{Leinweber Center for Theoretical Physics, Randall Laboratory of Physics \\The University of Michigan, Ann Arbor, MI 48109-1040, USA }

\emailAdd{junhoh@umich.edu}
\emailAdd{larsenf@umich.edu}
\emailAdd{jimliu@umich.edu}

\makeatletter
\newcommand*{\rom}[1]{\expandafter\@slowromancap\romannumeral #1@}
\makeatother

\abstract{
We study nearly extreme black holes with nearly AdS$_2$ horizon geometry in various settings inspired by string theory. 
Our focus is on the scales of the nAdS$_2$ region and their relation to microscopic theory. These scales are determined by a generalization of the 
attractor mechanism for extremal black holes and realized geometrically as the normal derivatives along the 
extremal attractor flow. In some cases the scales are equivalently determined by the charge dependence of the extremal attractor by itself. 
Our examples include near extreme black holes in $D\geq 4$ dimensions, AdS boundary conditions, rotation, and 5D black holes on the non-BPS branch. 
}

\begin{document}
\maketitle
\flushbottom

\section{Introduction}

The holographic correspondence between nearly AdS$_2$ geometry and nearly CFT$_1$'s \cite{Maldacena:2016hyu,Maldacena:2016upp,Kitaev:2017awl} is not conformal. Thus both sides of the duality depend on a scale and for both it is natural to focus on effective quantum field theory near the IR. There the relevant description for the two sides is Jackiw-Teitelboim gravity \cite{Teitelboim:1983ux,Jackiw:1984je}
(and its relatives \cite{Grumiller:2002nm,Castro:2018ffi,Moitra:2019bub,Castro:2019crn}) and the 
melonic theories (such as the SYK model \cite{Sachdev:1992fk,Kitaev,Rosenhaus:2018dtp}), respectively.

However, black hole physics is the key motivation for studying AdS$_2$ quantum gravity and the most confounding black hole puzzles demand a subtle interplay between the IR and the UV \cite{Mathur:2009hf,Harlow:2014yka}. It is therefore important to embed the nearly AdS$_2$ geometry into a UV complete 
setting (such as string theory) 
and analyze its IR features also from the UV point of view. In this paper we develop a simple and general mechanism that relates the IR parameters of effective quantum field theory near the IR to the underlying UV description. 

Our inspiration is the attractor mechanism for extremal black holes \cite{Ferrara:1996dd,Strominger:1996kf,Astefanesei:2006dd}. In that setting the parameters of the effective IR description depend only on conserved black hole charges and not on coupling constants (moduli) of the underlying UV theory. Moreover, precise relations between the IR and UV parameters are determined by attractor equations that are algebraic rather than differential. These features of the attractor mechanism facilitate quantitative comparisons between the IR and the UV in the extremal setting \cite{Sen:2007qy,Dabholkar:2014ema}. They are related to enhanced symmetry in the AdS$_2$ horizon region, such as the conformal symmetry that emerges there. 

The physical underpinning of the extremal attractor mechanism is radial evolution along the infinite distance to the extremal horizon. The near extreme black hole geometry is qualitatively different because the event horizon is located at a finite distance, so it is far from obvious that a generalization of the extremal attractor mechanism applies in this situation. However, the power of effective quantum field theory is that it organizes symmetries systematically, also when they are broken. This principle ensures an attractor mechanism also for near extremal black holes. We refer to it
as a nAttractor mechanism \cite{Larsen:2018iou}.

As illustration, consider a static, spherically symmetric black hole in $D$ spacetime dimensions. The area of the spheres surrounding the black hole take the form $R(r)^{D-2}\Omega_{D-2}$ where $R(r)$ is a monotonic function of the radial coordinate $r$ and $\Omega_{D-2}$ is the area of the unit sphere $S^{D-2}$. In this notation the area law for the black hole entropy is
\begin{equation}
	\label{eqn:Sintro}
	S_{\rm ext} = \frac{A}{4G} = \frac{R(r_0)^{D-2}\Omega_{D-2}}{4G}~,
\end{equation}
where $r=r_0$ locates the horizon in this coordinate system. The attractor mechanism demands that, for an extremal black hole, the value of the radial function at the horizon $R(r_0)$ depends on conserved charges but is independent of other asymptotic data (such as the asymptotic values of scalar fields, a.k.a. moduli). Moreover, the dependence of $R(r_0)$ on conserved charges can be determined by solving algebraic attractor equations or by extremizing the entropy function \cite{Larsen:2006xm,Sen:2007qy}. Therefore the dependence of the extremal black hole entropy \eqref{eqn:Sintro} on conserved charges can be computed without finding the black hole geometry. 

The {\it standard} version of a nAttractor mechanism generalizes these results away from extremality. The entropy of a near extremal black hole exceeds the extremal value by a term that is linear in the temperature and identified as the specific heat \cite{Preskill:1991tb,Almheiri:2016fws}. 
\begin{equation}
	\label{eqn:Lintr}
S - S_{\rm ext} = 	\left(\fft{\partial S}{\partial T}\right)_{T\to0} ~T =\left(\fft{C_{Q,P}}{T}\right)_{T\to0}~T~.
\end{equation}
The coefficient of the temperature is a length scale $L$ given by the nAttractor formula: \footnote{In \cite{Larsen:2018iou} a length scale was introduced as 
$L_{\rm there} = \fft12\pi L_{\rm here}$. The normalization was chosen so $L_{\rm there}$ coincides with the ``long string scale'' in the UV theory, at least in the simplest cases. The normalization $L_{\rm here}$ is advantageous presently because it generalizes to any dimension.}
\begin{equation}
	\label{eqn:Lintro}
	L= \left(\fft{C_{Q,P}}{T}\right)_{T\to0}=2\pi S_{\rm ext}  \left(\fft{dR^2}{dr}\right)_{r=r_0}~.
\end{equation}
The scale $L$ characterizes the excitations above the ground state, yet the function $R^2$ that is differentiated in a nAttractor formula is the one describing the {\it extremal} black hole. This is a major simplification because it is much more challenging to find a nonextremal black hole solution than an extremal one.

The radial derivative probes the extremal geometry infinitesimally {\it beyond} the AdS$_2$ horizon region. This is satisfying from an effective quantum field theory point of view because this simple device realizes geometrically the determination of the IR scale by {\it matching} with the UV data embodied in the attractor {\it flow} (radial dependence). Moreover, for a BPS black hole it is a desirable bonus that the entire flow preserves supersymmetry so the dependence of the low energy
scale $L$ on UV parameters enjoys some protection against renormalization. 

The scale $L$ (\ref{eqn:Lintro}) is the {\it conformal symmetry breaking scale} of the nAdS$_2$/CFT$_1$ correspondence. Thus $L$ is the dimensionful coupling of the famous Schwarzian boundary theory of nAdS$_2$ holography 
\cite{Kitaev,Maldacena:2016hyu,Maldacena:2016upp,Stanford:2017thb,Mertens:2017mtv}. In more general settings there are many other scales. For example, in the context of ${\cal N}=2$ supergravity, there is a complex scalar field $z^i$ in each vector multiplet. The extremal attractor mechanism determines the value of these scalars at the horizon in terms of black hole charges while the nAttractor mechanism gives a (complex) scale analogous to (\ref{eqn:Lintro}) for each scalar. These other scales are the dimensionful couplings entering the boundary theory describing vector fields in nAdS$_2$ \cite{Davison:2016ngz,Moitra:2018jqs,Sachdev:2019bjn}. 

There is a {\it strong} version of the nAttractor mechanism where the radial derivative in (\ref{eqn:Lintro}) is replaced by a gradient in the space of black hole charges. This realizes the intuition that a small motion normal to the horizon can be mimicked by adjusting the black hole charges such that the horizon itself moves. What makes this version of the nAttractor mechanism  much stronger is that the scale follows entirely from the analysis of the extremal attractor mechanism, it is a corollary that does not rely on the attractor flow. The catch is that the ``exchange rate'' between the geometric normal derivative and its analogue in charge space involves the scalar moduli quite nontrivially. As we discuss, this can present practical and conceptual obstacles in more elaborate examples. 

The main features of the nAttractor mechanism were established already in \cite{Larsen:2018iou}. Although details were only worked out for near extremal 
versions of 4D BPS black holes in ungauged ${\cal N}=2$ supergravity, similar results were anticipated in many other settings. The goal of this article is to confirm this expectation and provide the details needed for practical computations in various contexts. Accordingly, we will generalize the nAttractor mechanism to many other families of near-extremal black holes. For example, we consider higher dimensions, AdS, and examples with rotation. In all cases we find a standard nAttractor mechanism that gives the scale of the IR theory in terms of UV variables in a form analogous to (\ref{eqn:Lintro}), involving a normal derivative of the radial function.

However, the strong nAttractor mechanism, formulated as a derivative in charge space, proves more involved. The dependence on scalar moduli that enters the trade between a radial derivative in spacetime and a gradient in the space of charges is not entirely universal; the details are somewhat model dependent. The simplest version of the strong nAttractor is predicated on a one-to-one map between ratios of conserved charges and attractor moduli. This is indeed the structure of the prototypical attractor mechanism for BPS black holes in 4D ungauged ${\cal N}=2$ supergravity where all matter is in vector-multiplets. Some departures from this baseline setting, such as the addition of hyper-multiplets, can be addressed by cosmetic changes. However, other charges, such as the non-BPS branch, involve more model dependence. We explain the general situation and work out all details in an illustrative family of models. Additional discussion of this example are left for our discussion in section \ref{discussion}. 

This paper is organized as follows. In section \ref{nAttractor} we introduce a nAttractor mechanism in a simple 4D setting. Section \ref{standard} is divided into several subsections that each provide a generalization of the standard nAttractor. We consider in turn higher dimensions $D\geq 4$, AdS, and rotation. In each of these subsections we work out an explicit example. 
In section \ref{strong} we develop the generalization of the strong nAttractor. We focus on the $ST(N)$ model in 5D since this example illustrates the role of flat directions well. Section \ref{discussion} summarized our results and discuss promising avenues for future directions.

\section{A nAttractor Mechanism}
\label{nAttractor}
In this section we briefly review the nAttractor mechanism, largely following \cite{Larsen:2018iou}. This discussion will serve as reference when considering more elaborate generalizations in the remainder of the paper. 

\subsection{A 4D Setting}
\label{nAttractor:setting}
The starting point is 4D Einstein gravity coupled to an arbitrary number of complex scalar fields $z^i$ (with $i,j=1,2,\cdots,n$) and vector fields $A^I$ (with $I,J=0,1,\cdots, N$) through 
the action
\begin{equation}
	S=\fft{1}{16\pi G_4}\int\left[R^{(4)}(*\mathbbm{1})-2G_{i\bar j}(*dz^i)\wedge d\bar z^j-\fft{1}{2}\mu_{IJ}(*F^I)\wedge F^J-\fft{1}{2}\nu_{IJ}F^I\wedge F^J\right]~.\label{4D:action}
\end{equation}
The kinetic functions $\mu_{IJ}$, $\nu_{IJ}$, and $G_{i\bar j}$ generally depend on the scalar fields $z^i$. The bosonic part of ${\cal N}\geq 2$ ungauged supergravity coupled to matter fields in vector and hyper-multiplets is a large class of theories with actions of this form. A few aspects of this theory are reviewed in Appendix \ref{App:A}.

The most general static, spherically symmetric black hole solution to these theories takes the form \cite{Ferrara:1997tw}
\begin{subequations}
\begin{align}
	ds^2&=-\fft{r^2-r_0^2}{R^2}dt^2+\fft{R^2}{r^2-r_0^2}dr^2+R^2d\Omega_{2}^2~,\label{4D:metric}\\
	P^I\,\mathrm{vol}(S^2)&=F^I|_{S^2}~,\\
	Q_I\,\mathrm{vol}(S^2)&=\mu_{IJ}(*F^J)|_{S^2}+\nu_{IJ}(F^J)|_{S^2}~.
\end{align}\label{4D:ansatz}%
\end{subequations}
The radial function $R$ and the scalar fields $z^i$ (entering through $\mu_{IJ}, \nu_{IJ}$) are functions of the radial coordinate $r$ as well as parameters specifying the black hole data. We will not always make all these variables explicit but we discuss them in detail below. 

The \emph{ansatz} above automatically satisfies the Bianchi identity and the equation of motion for the vector fields $A^I$. The Einstein equations and the scalar equation of motion that determine $R, z^i$ are equivalent to a simple mechanical problem where ``particles'' with coordinates $R, z^i$ move in ``time'' $r$ under the influence of an effective potential 
\begin{align}
	V_{\mathrm{eff}}&=V_{\mathrm{eff}}(\mu_{IJ},\nu_{IJ},P^I,Q_I)\nn\\&=
	\begin{pmatrix}
	P^I & Q_I
	\end{pmatrix}
	\begin{pmatrix}
	\mu_{IJ}+\nu_{IK}(\mu^{-1})^{KL}\nu_{LJ} & -\nu_{IK}(\mu^{-1})^{KJ} \\
	-(\mu^{-1})^{IK}\nu_{KJ} & (\mu^{-1})^{IJ}
	\end{pmatrix}
	\begin{pmatrix}
	P^J \\ Q_J
	\end{pmatrix}~.\label{4D:eff:potential}
\end{align}
In this section it will be sufficient to rely on the intuition embodied in the mechanical analogue, rather than the explicit equations of motion. 

As we have already mentioned, $R, z^i$ are not just functions of the radial coordinate $r$; they also depend on black hole data: the mass $M$, conserved dyonic charges $(P^I,Q_I)$, and the asymptotic values of scalar fields $z^i_\infty$. The position of the event horizon $r_0$ is also a function of all black hole data and it is convenient to eliminate explicit dependence on $M$ in favor of the variable $r_0$. For example, consider the extremal mass $M_{\rm ext}$, the lower bound for the black hole mass with given conserved charges and asymptotic values of scalar fields. It can be quite complicated to compute $M_{\rm ext}$ as function of the other black hole parameters and it may anyway by unilluminating. In contrast, in terms of the geometrical parameter $r_0$ the extremal limit is clearly $r_0=0$, because that is the value of $r_0$ where the coefficients in the metric \eqref{4D:metric} develop double poles. After this change of variables, we 
write the functions $R$ and $z^i$ as
\begin{subequations}
\begin{align}
	R&=R(r;r_0,P^I,Q_I,z^i_\infty)~,\label{4D:radial}\\
	z^i&=z^i(r;r_0,P^I,Q_I,z^i_\infty)~.
\end{align}
\end{subequations}
We assume that these $R, z^i$ are smooth with respect to both $r$ and $r_0$ for $r\geq r_0$. 

We can compute the black hole entropy and the temperature for any metric of the form (\ref{4D:metric}). They are
\begin{align}
	S=S(r_0,P^I,Q_I,z^i_\infty)&=\fft{\pi}{G_4} R(r_0;r_0,P^I,Q_I,z^i_\infty)^2~,\label{4D:entropy}\\ 
	T=T(r_0,P^I,Q_I,z^i_\infty)&=\fft{r_0}{2\pi R(r_0;r_0,P^I,Q_I,z^i_\infty)^2}~.\label{4D:temp}
\end{align}
We are interested in black holes with strictly positive entropy so (\ref{4D:entropy}) demands $R>0$. 
Then the temperature (\ref{4D:temp}) shows that the extremal limit $T\to 0$ is equivalent to $r_0\to0$. We see again that the horizon position $r_0$ parametrizes the departure from the extremal limit and we refer to $r_0$ as an extremality parameter when we have this aspect in mind.

We stress that the entropy (\ref{4D:entropy}) and the temperature (\ref{4D:temp}) depend on the extremality parameter $r_0$ for two distinct reasons. First, they depend on $r_0$ as a parameter that characterizes a given black hole solution, denoted by the 2nd argument of the radial function $R(r_0;r_0,P^I,Q_I,z^i_\infty)$ (after the semicolon). On the other hand, they also depend on $r_0$ as a location where the radial function is evaluated, denoted by the 1st argument of the radial function $R(r_0;r_0,P^I,Q_I,z^i_\infty)$ (before the semicolon). This point is essential for our considerations so, as a visual reminder, we have introduced the semicolon to distinguish the radial coordinate and the black hole parameters.

\subsection{The Extremal Attractor Mechanism}
The original attractor mechanism applies to extremal black holes ($r_0=0$). In its simplest form, it posits that the entropy and the horizon values of scalar fields 
are determined by extremizing the effective scalar potential \eqref{4D:eff:potential} $V_{\mathrm{eff}}$ with respect to the scalar fields $z^i$. Thus
\begin{align}
	S_{\mathrm{ext}}&=S(0,P^I,Q_I,z^i_\infty)=\fft{\pi}{4G_4} V_{\mathrm{eff}}~\mbox{at~the~critical~points}~,\label{4D:att1}\\
	z_{\mathrm{ext}}^i|_{\mathrm{hor}}&=z^i(0;0,P^I,Q_I,z^i_\infty)=\mbox{critical~points~of~}V_{\mathrm{eff}}~.\label{4D:att2}
\end{align}
The notation $z_{\mathrm{ext}}^i|_{\mathrm{hor}}$ stresses again that the two zeros in $z^i(0;0,P^I,Q_I,z^i_\infty)$ are distinct. The first indicates the location (the horizon) and the second refers to the parameters of the black hole (extremal). 

The power of the extremal attractor mechanism is that it allows the computation of some horizon data ($S_{\mathrm{ext}}$ and $z^i_{\mathrm{ext}}|_{\mathrm{hor}}$) by solving algebraic equations (extremizing a potential) rather than from the complete solution to the equations of motion which are differential. In some examples all moduli $z^i$ are fixed by the extremization. In the context of 4D $\mathcal N=2$ ungauged supergravity, an example of this situation is BPS black holes when all matter is in vector multiplets. More generally, some moduli may decouple from the dynamics and their values are arbitrary even after extremization. Hypermultiplets is a simple example in the same context. The flat directions on the non-BPS branch is another example. It is important to distinguish moduli that are fixed by 
the attractor mechanism and those that are not. We will encounter both types of moduli in our examples. 

\subsection{A nAttractor Mechanism for Near-Extremal Black Holes}
\label{nAttractor:nAttractor}
The goal of the nAttractor mechanism is to generalize the extremal attractor mechanism to \emph{near}-extremal black holes. It determines the entropy and the scalar fields in the horizon region of a near-extremal black hole without referring to the explicit black hole solution. 

The entropy of a \emph{near}-extremal black hole is linear in the (small) temperature \cite{Preskill:1991tb,Almheiri:2016fws}: 
\begin{align}
	S&=S_{\mathrm{ext}}+T\lim_{T\to0}\left(\fft{\partial S(r_0,P^I,Q_I,z^i_\infty)}{\partial T}\right)_{P^I,Q_I,z^i_\infty}+\mathcal O(T^2)~.\label{nEXT}
\end{align}
Since $S_{\mathrm{ext}}$ is determined by the extremal attractor mechanism and the entropy is related to the radial function $R$ through (\ref{4D:entropy}), we can focus on the slope of the linear function for the entropy. It is encoded in 
\begin{align}
	L&\equiv\fft{\pi}{G_4}\lim_{T\to0}\left(\fft{\partial R(r_0;r_0,P^I,Q_I,z^i_\infty)^2}{\partial T}\right)_{P^I,Q_I,z^i_\infty}~.\label{length:scale}
\end{align}
This length scale can be interpreted physically as the heat capacity per thermal unit of the excitations above the ground state \cite{Almheiri:2016fws}
\begin{equation}
	L = \left(\fft{C_{Q,P}}{T}\right)_{T\to 0}~.
\end{equation}

The nAttractor mechanism involves two simplifications in computing the length scale $L$. First, the derivative with respect to temperature in (\ref{length:scale}) can be traded for a derivative with respect to the geometrical parameter $r_0$. This is due to the relation (\ref{4D:temp}) for the temperature. Second, the resulting radial derivative can be \emph{computed in the background of the corresponding extremal black holes} \cite{Larsen:2018iou}. Taking these simplifications together, the length scale (\ref{length:scale}) becomes
\begin{align}
	L&=2\pi S_{\mathrm{ext}}\lim_{r\to0}\left(\fft{\partial R(r;0,P^I,Q_I)^2}{\partial r}\right)_{P^I,Q_I}~.\label{4D:STEP1}
\end{align}
The nAttractor aspect of these simplifications is that the geometry of the nearly extremal black hole has been eliminated altogether. It is sufficient to consider the geometry of the corresponding 
extremal black hole. Such solutions are much simpler because they involve differential equations of first order rather than of second order. 
We will refer to this form of a nAttractor as the \emph{standard} nAttractor mechanism. 

In some settings we can further simplify (\ref{4D:STEP1}) by rewriting the radial derivative as a \emph{gradient with respect to conserved charges}. Generally there are many conserved charges so the resulting gradient must be contracted with a ``normal'' vector $(p^I_\infty,q_I^\infty)$ in the space of charges:
\begin{align}
	L&=2\pi S_{\mathrm{ext}}\sum_I\left[q^\infty_I\left(\fft{\partial R(0;0,P^I,Q_I)^2}{\partial Q_I}\right)_{P^I,Q_I}+p_\infty^I\left(\fft{\partial R(0;0,P^I,Q_I)^2}{\partial Q_I}\right)_{P^I,Q_I}\right]\label{N=2:4D:STEP2}~.
\end{align}
The ``normal'' vector $(p^I_\infty,q_I^\infty)$ encodes the dependence on asymptotic scalars $z^i_\infty$ \cite{Larsen:2018iou}. 
The situation is the inverse of the extremal attractor mechanism. There the horizon scalars $z^i_{\mathrm{ext}}|_{\mathrm{hor}}=z^i(0;0,P^I,Q_I)$ are 
determined by the extremization principle (\ref{4D:att2}) for any input charge vector $(Q_I, P^I)$. In the nAttractor mechanism, the asymptotic scalars $z^i_{\mathrm{ext}}|_{\mathrm{asympt}}$ are given inputs that specify the ``normal'' vector $(p^I_\infty,q_I^\infty)$ in charge space as   
\begin{equation}
	z^i_{\mathrm{ext}}|_{\mathrm{asympt}}=z^i(0;0,p^I_\infty,q_I^\infty)~.\label{normal:asymptotic}
\end{equation}
In other words, $(p^I_\infty,q_I^\infty)$ is the vector of conserved charges that would give $z^i_{\mathrm{ext}}|_{\mathrm{asympt}}$ as attractor values. 
The ``normal''  vector specified this way is not entirely unique but this ambiguity can be addressed by imposing additional conditions. 

The formula for the scale \eqref{N=2:4D:STEP2} is truly remarkable. It relies entirely on input from the {\it extremal} attractor mechanism: the entropy and 
horizon scalars as functions of conserved charges. We will refer to the form (\ref{N=2:4D:STEP2}) of a nAttractor as the \emph{strong} nAttractor mechanism. 

We must caution that the strong nAttractor mechanism have been established only in some favorable cases, such as the BPS black holes in 4D $\mathcal N=2$ ungauged supergravity \cite{Larsen:2018iou}. A delicate point is the algorithm encoding the asymptotic scalars $z^i_\infty$ (\ref{normal:asymptotic})
in the ``normal'' vector. For example, due care must be taken to account for the fact that generally only some scalars are fixed by the extremal attractor mechanism; others remain as moduli. The complications inherent in the definition of $(p^I_\infty,q_I^\infty)$ present obstacles that we have not been able to address in all settings.

In this section we have for simplicity focused on the length scale $L$ associated to the entropy of a \emph{near}-extremal black hole, which characterizes the approach to the black hole horizon of the radial function $R$. There are analogous length scales $L^i$ associated to the horizon scalars of a \emph{near}-extremal black hole, which quantifies the corresponding behavior of the scalar fields $z^i$. They can be obtained analogously, by replacing $R^2$ with $\fft{1}{\pi}G_4z^i$ in (\ref{length:scale}~, \ref{4D:STEP1}~, \ref{N=2:4D:STEP2}). 

In the following sections, we derive detailed versions of the nAttractor formulae (\ref{4D:STEP1}, \ref{N=2:4D:STEP2}) that are applicable in a various specific settings that were not previously considered in the literature.

\section{Near Extremal Black Hole Thermodynamics from Radial Derivatives}
\label{standard}

In this section we generalize the standard nAttractor mechanism from spherically symmetric 4D black holes to three other settings: higher dimensional black holes, black holes in AdS, and rotating black holes. In each case we first define the length scales associated to the entropy and the horizon scalars of a near-extremal black hole following (\ref{nEXT}--\ref{length:scale}) and then we show how they can be computed from radial derivatives along the attractor flow of the corresponding extremal black hole solution by adapting subsection \ref{nAttractor:nAttractor}.

\subsection{Higher Dimensional Black Holes}

\label{D-dim}
We consider a $D$-dimensional Einstein-Hilbert action coupled to arbitrary number of real scalar fields $\phi^i$ (with $i,j=1,2,\cdots,n$) and vector fields $A^I$ (with $I,J=0,1,\cdots, N$),
\begin{equation}
	S=\fft{1}{16\pi G_D}\int\left[R^{(D)}(*\mathbbm{1})-\fft{1}{2}G_{ij}[\phi](*d\phi^i)\wedge d\phi^j-\fft{1}{2}\mu_{IJ}[\phi](*F^I)\wedge F^J\right]~.\label{D-dim:action}
\end{equation}
We analyze a general static, spherically symmetric, electrically charged black hole solution to this theory. Without loss of generality it can be written as
\begin{subequations}
	\begin{align}
	ds^2&=-e^{2\Phi}dt^2+e^{-2\Phi}dr^2+R^2d\Omega^2_{D-2}~,\label{D-dim:ansatz:metric}\\
	Q_I\,\mathrm{vol}(S^{D-2})&=\mu_{IJ}*F^J~.\label{D-dim:ansatz:vector}
	\end{align}\label{D-dim:ansatz}%
\end{subequations}
The metric functions $\Phi$, $R$ and the scalar fields $\phi^i$ depend on the radial coordinate $r$ as well as the black hole data: mass $M$, conserved electric charges $Q_I$, and asymptotic values of scalar fields $\phi^i_\infty$. 

The ansatz (\ref{D-dim:ansatz}) automatically satisfies the Bianchi identity and the equation of motion for the vector fields $A^I$
\begin{equation}
	dF^I=0\qquad\mbox{and}\qquad d(\mu_{IJ}*F^J)=0~.\label{Bianchi:eom}
\end{equation}
The scalar equation of motion is given as
\begin{equation}
	0=R^{2-D}\partial_r(G_{ij}R^{D-2}e^{2\Phi}\partial_r\phi^j)-\fft{1}{2}e^{2\Phi}(\partial_iG_{jk})\partial_r\phi^j\partial_r\phi^k-\fft{1}{2}\fft{\partial_iV_{\mathrm{eff}}}{R^{2(D-2)}}~,\label{D-dim:scalar:eom}
\end{equation}
where we have defined the effective potential as
\begin{equation}
	V_{\mathrm{eff}}=V_{\mathrm{eff}}(\mu_{IJ},Q_I)=(\mu^{-1})^{IJ}Q_IQ_J~.
\end{equation}
Finally, we have three independent Einstein equations,
\begin{subequations}
\begin{align}
	0&=R^{4-D}\partial_r(R^{4-D}\partial_r(R^{2(D-3)}e^{2\Phi}))-2(D-3)^2~,\label{D-dim:Einstein:1}\\
	0&=R^{4-D}\partial_r(e^{2\Phi}R^{D-3}\partial_rR)-(D-3)+\fft{1}{2(D-2)}\fft{V_{\mathrm{eff}}}{R^{2(D-3)}}~,\label{D-dim:Einstein:2}\\
	0&=(D-2)\fft{\partial_r^2R}{R}+\fft{1}{2}G_{ij}\partial_r\phi^i\partial_r\phi^j~.\label{D-dim:Einstein:3}
\end{align}\label{D-dim:Einstein}%
\end{subequations}
They are $E_{00}-(D-3)E_{22}$, $E_{22}$, and $E_{00}+E_{11}$ respectively, where $\{e^0,e^1,\cdots,e^{D-1}\}$ denotes the coframe of the metric (\ref{D-dim:ansatz:metric}). 

We can rewrite the equations of motion compactly in terms of a new radial coordinate $\rho$ defined as $\rho=(D-3)\int R^{D-4}dr$. In particular, the first Einstein equation (\ref{D-dim:Einstein:1}) is simplified dramatically in terms of $\rho$ and therefore can be solved for $e^{2\Phi}$ as
\begin{equation}
	0=\partial_\rho^2(R^{2(D-3)e^{2\Phi}})-2\quad\to\quad e^{2\Phi}=\fft{\rho^2-\rho_0^2}{R^{2(D-3)}}~,\label{D-dim:new:radial}
\end{equation}
where the integration constant in $\rho=(D-3)\int R^{D-4}dr$ is chosen to remove a possible linear term in the numerator of the expression for $e^{2\Phi}$. Substituting (\ref{D-dim:new:radial}) into the metric ansatz (\ref{D-dim:ansatz:metric}) gives
\begin{equation}
	ds^2=-\fft{\rho^2-\rho_0^2}{R^{2(D-3)}}dt^2+\fft{R^2}{(D-3)^2(\rho^2-\rho_0^2)}d\rho^2+R^2d\Omega^2_{D-2}~,\label{D-dim:BH}
\end{equation}
and it becomes clear that the parameter $\rho_0$ indicates the horizon position in these coordinates. We rewrite the scalar equation of motion (\ref{D-dim:scalar:eom}) and the remaining Einstein equations (\ref{D-dim:Einstein:2}--\ref{D-dim:Einstein:3}) using the black hole metric (\ref{D-dim:BH}) as
\begin{subequations}
\begin{align}
	0=&~\partial_\rho((\rho^2-\rho_0^2)G_{ij}\partial_\rho\phi^j)-\fft{1}{2}(\rho^2-\rho_0^2)(\partial_iG_{jk})\partial_\rho\phi^j\partial_\rho\phi^k-\fft{1}{2(D-3)^2}\fft{\partial_iV_{\mathrm{eff}}}{R^{2(D-3)}}~,\label{D-dim:eom:phi}\\
	0=&~\partial_\rho((\rho^2-\rho_0^2)\partial_\rho\log R)-\fft{1}{D-3}+\fft{1}{2(D-2)(D-3)^2}\fft{V_{\mathrm{eff}}}{R^{2(D-3)}}~,\label{D-dim:eom:R}\\
	0=&~\fft{\partial_\rho^2 R}{R}+(D-4)\fft{(\partial_\rho R)^2}{R^2}+\fft{1}{2(D-2)}G_{ij}\partial_\rho\phi^i\partial_\rho\phi^j~.
\end{align}\label{D-dim:eom}%
\end{subequations}
Summarizing so far, we found the general metric (\ref{D-dim:BH}) for static, spherically symmetric, electrically charged $D$-dimensional black hole solutions. 
The radial function $R$ and the scalar fields $\phi^i$ satisfy the equations of motion (\ref{D-dim:eom}). 

As in section \ref{nAttractor:setting}, it is advantageous to use the horizon position $\rho_0$ as a variable characterizing the black hole, effectively 
replacing the black hole mass $M$. We
therefore write the radial function $R$ and the scalar fields $\phi^i$ as
\begin{subequations}
\begin{align}
	R&=R(\rho;\rho_0,Q_I,\phi^i_\infty)~,\label{D-dim:radial}\\
	\phi^i&=\phi^i(\rho;\rho_0,Q_I,\phi^i_\infty)~.
\end{align}
\end{subequations}
We assume that $R,\phi^i$ are smooth with respect to both $\rho$ and $\rho_0$ outside the event horizon $\rho\geq\rho_0$.

We can compute the entropy and the temperature of the black hole (\ref{D-dim:BH}) in terms of the radial function (\ref{D-dim:radial}) as
\begin{align}
	S(\rho_0,Q_I,\phi^i_\infty)&=\fft{\Omega_{D-2}}{4G_D}R(\rho_0;\rho_0,Q_I,\phi^i_\infty)^{D-2}~,\label{D-dim:entropy}\\
	T(\rho_0,Q_I,\phi^i_\infty)&=\fft{(D-3)\rho_0}{2\pi R(\rho_0;\rho_0,Q_I,\phi^i_\infty)^{D-2}}~.\label{D-dim:temp}
\end{align}
Regular black holes have strictly positive entropy so (\ref{D-dim:entropy}) shows $R(\rho_0;\rho_0,Q_I,\phi^i_\infty)>0$. 
From (\ref{D-dim:temp}) we then see that the extremal limit $T\to 0$ is equivalent to $\rho_0\to0$. Thus the parameter $\rho_0$ is a measure of the distance from
extremality. The entropy of an extremal black hole, which we abbreviate as $S_{\mathrm{ext}}$, is therefore given as $S_{\mathrm{ext}}=S(0,Q_I,\phi^i_\infty)$.

\subsubsection{The Standard nAttractor Mechanism}
\label{D-dim:nAttractor}
As discussed in section \ref{nAttractor:nAttractor}, the near-extremal black hole entropy is conveniently encoded in the slope of the heat capacity at vanishing temperature. Following (\ref{nEXT}--\ref{length:scale}) we therefore define the length scale characterizing the entropy of a higher dimensional near-extremal 
black hole as
\begin{align}
	L&\equiv\lim_{T\to0}\left(\fft{\partial S}{\partial T}\right)_{Q_I}=\fft{\Omega_{D-2}}{4G_D}\lim_{T\to0}\left(\fft{\partial R(\rho_0;\rho_0,Q_I,\phi^i_\infty)^{D-2}}{\partial T}\right)_{Q_I,\phi^i_\infty}~,\label{D-dim:length:scale}
\end{align}
and we seek to compute it using the standard nAttractor mechanism.

We first trade the temperature derivative for differentiation with respect to the horizon position $\rho_0$ and then 
substitute (\ref{D-dim:entropy}--\ref{D-dim:temp}) into the result:
\begin{align}
	L&=\lim_{\rho_0\to0}\left(\fft{\partial T(\rho_0,Q_I,\phi^i_\infty)}{\partial\rho_0}\right)^{-1}_{Q_I,\phi^i_\infty}\left(\fft{\partial S(\rho_0,Q_I,\phi^i_\infty)}{\partial\rho_0}\right)_{Q_I,\phi^i_\infty}\nn\\&=\fft{2\pi S_{\mathrm{ext}}}{D-3}\lim_{\rho_0\to0}\left(\fft{\partial R(\rho_0;\rho_0,Q_I,\phi^i_\infty)^{D-2}}{\partial\rho_0}\right)_{Q_I,\phi^i_\infty}~.\label{D-dim:length:scale:1}
\end{align}
Since the $\rho_0$-derivative acts on both the 1st and the 2nd argument of the radial function $R(\rho_0;\rho_0,Q_I,\phi^i_\infty)$ this formula still requires the full non-extremal solution as an input. Generally it therefore still demands quite elaborate computations. However, we next simplify (\ref{D-dim:length:scale:1}) further as
\begin{align}
	L&=\fft{2\pi S_{\mathrm{ext}}}{D-3}\lim_{\rho=\rho_0\to0}\left[\left(\fft{\partial R(\rho;\rho_0,Q_I,\phi^i_\infty)^{D-2}}{\partial\rho}\right)_{\rho_0,Q_I,\phi^i_\infty}+\left(\fft{\partial R(\rho;\rho_0,Q_I,\phi^i_\infty)^{D-2}}{\partial\rho_0}\right)_{\rho,Q_I,\phi^i_\infty}\right]\nn\\
	&=\fft{2\pi S_{\mathrm{ext}}}{D-3}\lim_{\rho\to0}\left(\fft{\partial R(\rho;0,Q_I,\phi^i_\infty)^{D-2}}{\partial\rho}\right)_{Q_I,\phi^i_\infty}~.\label{D-dim:length:scale:2}
\end{align}
In order to get the 2nd equation, we use the condition \footnote{In \cite{Larsen:2018iou}, the condition (\ref{D-dim:regularity}) corresponds to the finiteness of $\partial_MR^2$ at extremality $M\to M_{\mathrm{ext}}$. Similar conditions in the following subsections such as (\ref{Cosmo:regularity}) and (\ref{Rot:regularity}) can be understood analogously.}
\begin{equation}
	R(\rho;\rho_0,Q_I,\phi^i_\infty)=R(\rho;0,Q_I,\phi^i_\infty)+\mathcal O(\rho_0^2).\label{D-dim:regularity}
\end{equation}
This condition is satisfied for any radial function $R(\rho;\rho_0,Q_I,\phi^i_\infty)$ that is obtained perturbatively with respect to $\rho_0^2$ from the extremal ($\rho_0=0$) solutions to (\ref{D-dim:eom}). This is not a strong assumption since (\ref{D-dim:eom}) depends on $\rho_0$ through $\rho_0^2$ only. 

Physically, the regularity condition \eqref{D-dim:regularity} states that the final approach to extremality is dominated by the position of the horizon, rather than the dependence of the metric on the black hole mass. As a result, the length scale defined in (\ref{D-dim:length:scale}) can be obtained from the radial derivative of $R$ in the background of the corresponding {\it extremal} black hole (\ref{D-dim:length:scale:2}); there is no need for the geometry of a non-extremal black hole. This is the central statement of the standard nAttractor mechanism for a higher dimensional near-extremal black hole. 

In this subsection we focused on the length scale $L$ quantifying the entropy of a higher dimensional near-extremal black hole. The length scales $L^i$ that characterize the final approach to the horizon of the scalars $\phi^i$ can be obtained analogously, by replacing $R^{D-2}$ 
with $4G_D\phi^i/\Omega_{D-2}$ in (\ref{D-dim:length:scale}, \ref{D-dim:length:scale:1}, \ref{D-dim:length:scale:2}).

\subsubsection{Example: Black Holes in 5D \texorpdfstring{$\mathcal N=2$}{N=2} Ungauged Supergravity}\label{sec:D5N2sugra}

The action of 5D $\mathcal N=2$ ungauged supergravity coupled to $N$ vector multiplets is given as
\begin{equation}
	S=\int\left[R^{(5)}(*\mathbbm{1})-\fft{1}{2}G_{IJ}\left(*dX^I\wedge dX^J+*F^I\wedge F^J\right)-\fft16C_{IJK}A^I\wedge F^J\wedge F^K\right]~,\label{5D:sugra:action}
\end{equation}
where $I,J,K=0,1,\ldots,N$. The real scalar fields $X^I$ are constrained by $\mathcal V=1$ where $\mathcal V$ is the cubic prepotential
\begin{equation}
	\mathcal V=\fft16C_{IJK}X^IX^JX^K~,\label{5D:sugra:prepotential}
\end{equation}
with the structure constants $C_{IJK}$ completely symmetric on $IJK$.  The $N+1$ scalars $X^I$ subject to the constraint $\mathcal V=1$ can be parametrized by $N$ unconstrained scalar fields $\phi^i$ where $i,j,k=1,2,\ldots,N$.  The kinetic function $G_{IJ}$ is given as
\begin{equation}
	G_{IJ}=-\partial_I\partial_J\log\mathcal V|_{\mathcal V=1}=9X_IX_J-C_{IJK}X^K~,\label{5D:sugra:metric}
\end{equation}
where we have defined $X_I$ with lower index as
\begin{equation}
	X_I=\fft16C_{IJK}X^JX^K~.\label{5D:sugra:lower}
\end{equation}

The $\mathcal N=2$ supergravity action (\ref{5D:sugra:action}) can be obtained from the general $D$-dimensional action (\ref{D-dim:action}) with $D=5$ under the substitution
\begin{align}
	G_{ij}\to G_{IJ}\partial_{\phi^i}X^I\partial_{\phi^j}X^J,\qquad \mu_{IJ}\to G_{IJ},
\end{align}
and with the addition of a Chern-Simons term.  However, the Chern-Simons term in (\ref{5D:sugra:action}) does not contribute to any equations of motion for the static spherically symmetric electric black holes under consideration.  As a result, the general $D$-dimensional analysis developed earlier in this subsection applies, and hence the standard nAttractor mechanism holds for these near-extremal black holes in 5D $\mathcal N=2$ ungauged supergravity.

It is nevertheless worthwhile to present specific details for the black holes in $\mathcal N=2$ supergravity (\ref{5D:sugra:action}).  As in (\ref{D-dim:ansatz:vector}) and (\ref{D-dim:BH}), we take the metric and gauge field ansatz
\begin{subequations}
	\begin{align}
	ds^2&=-\fft{\rho^2-\rho_0^2}{R^4}dt^2+\fft{R^2}{4(\rho^2-\rho_0^2)}d\rho^2+R^2d\Omega^2_3~,\\
	G_{IJ}*F^J&=Q_I\,\mathrm{vol}(S^3)~,
	\end{align}\label{5D:N=2:ansatz}%
\end{subequations}
where the function $R(\rho)$ and the scalar fields $X^I(\rho)$ are functions of the radial coordinate $\rho$ and 
further depend on the black hole data: the horizon position $\rho_0$ (in lieu of the black hole mass $M$), conserved electric charges $Q_I$, and asymptotic values of the scalar fields $X^I_\infty$.  Given this ansatz, the scalar and Einstein equations reduce to
\begin{subequations}
	\begin{align}
	0=&~\partial_\rho((\rho^2-\rho_0^2)G_{IJ}\partial_\rho X^J)-\fft{1}{2}(\rho^2-\rho_0^2)(\partial_IG_{JK})\partial_\rho X^J\partial_\rho X^K-\fft{1}{8}\fft{\partial_I(V_{\mathrm{eff}}+\lambda\mathcal V)}{R^4}~,\label{5D:sugra:eom:scalar}\\
	0=&~\partial_\rho((\rho^2-\rho_0^2)\partial_\rho\log R)-\fft{1}{2}+\fft{1}{24}\fft{V_{\mathrm{eff}}}{R^4}~,\label{5D:sugra:eom:Einstein:1}\\
	0=&~\fft{\partial_\rho^2 R}{R}+\fft{(\partial_\rho R)^2}{R^2}+\fft{1}{6}G_{IJ}\partial_\rho X^I\partial_\rho X^J~,\label{5D:sugra:eom:Einstein:2}
	\end{align}\label{5D:sugra:eom}%
\end{subequations}
while the Maxwell equations are automatically satisfied.  Here the effective scalar potential takes the form
\begin{align}
	V_{\mathrm{eff}}=(G^{-1})^{IJ}Q_IQ_J~,\label{5D:sugra:Veff}
\end{align}
and $\lambda$ is introduced as a Lagrange multiplier that enforces the constraint $\mathcal V=1$.  The above system of equations is in fact equivalent to (\ref{D-dim:eom}) with $D=5$ and the original scalar fields $\phi^i$ rewritten in terms of the constrained scalars $X^I$.

\subsubsection{Detailed Example: 5D STU Black Holes}

As an example of explicit computation of the nAttractor length scales we focus on the STU model.  The five-dimensional STU model consists of two vector multiplets coupled to supergravity with $C_{123}=1$. Thus we have $\mathcal V=X^1X^2X^3$ and substitution into (\ref{5D:sugra:metric}) yields $G_{IJ}=(X^I)^{-2}\delta_{IJ}$.

The non-extremal 5D STU black hole solutions of this theory are given by \cite{Behrndt:1998jd}
\begin{subequations}
\begin{align}
	ds^2&=-\fft{\rho^2-\rho_0^2}{(H^1H^2H^3)^\fft23}dt^2+(H^1H^2H^3)^\fft13\left[\fft{d\rho^2}{4(\rho^2-\rho_0^2)}+d\Omega_3^2\right]~,\label{eq:STUg}\\
	X^I&=X^I_\infty\fft{(H^1H^2H^3)^\fft13}{H^I}~,\\
    *F^I&=(X^I)^{2}Q_I\,\mathrm{vol}(S^3)\qquad(\mbox{no sum on $I$})~,
\end{align}\label{5D:STU:BH}%
\end{subequations}
where $H^I$ are linear functions of $\rho$
\begin{equation}
	 H^I(\rho;\rho_0,Q_I,X^I_\infty)=\rho+\sqrt{\fft14(Q_IX^I_\infty)^2+\rho_0^2}\qquad(\mbox{no sum on $I$)}~.\label{H:def}
\end{equation}
Comparison of (\ref{eq:STUg}) and (\ref{D-dim:BH}) identifies the radial function 
\begin{equation}
	R(\rho;\rho_0,Q_I,X^I_\infty)=\prod_{I=1}^3H^I(\rho;\rho_0,Q_I,X^I_\infty)^\fft16~.\label{5D:STU:radial}
\end{equation}

At this point it is clear that the standard nAttractor mechanism holds since the radial function (\ref{5D:STU:radial}) is smooth with respect to both $\rho$ and $\rho_0$ for $\rho\geq\rho_0$ and also satisfies the condition (\ref{D-dim:regularity}). However, in this example we want to spell out all details explicitly. We can compute the entropy and the temperature of the black hole by substituting the radial function (\ref{5D:STU:radial}) into (\ref{D-dim:entropy}) and (\ref{D-dim:temp}) as
\begin{align}
	S(\rho_0,Q_I,X^I_\infty)&=\fft{\pi^2}{2G_5}\prod_{I=1}^3H^I(\rho_0;\rho_0,Q_I,X^I_\infty)^\fft12~,\label{5D:STU:entropy}\\
	T(\rho_0,Q_I,X^I_\infty)&=\fft{\rho_0}{\pi}\prod_{I=1}^3H^I(\rho_0;\rho_0,Q_I,X^I_\infty)^{-\fft12}~.\label{5D:STU:temp}
\end{align}
The entropy (\ref{5D:STU:entropy}) is strictly positive even in the strict extremal limit $\rho_0=0$, provided that $Q_1Q_2Q_3\neq0$. With the same proviso, the extremal limit $T\to 0$ is equivalent to $\rho_0\to0$, so the horizon position $\rho_0$ is an ``extremality parameter''. The entropy of the extremal black hole is therefore given as 
\begin{equation}
S_{\mathrm{ext}}=S(0,Q_I,X^I_\infty) = {\fft{\pi^2}{2G_5}} \sqrt{Q_1 Q_2 Q_3} ~.
\end{equation}

The length scale associated to the additional entropy of the near-extremal 5D $STU$  black hole, above and beyond $S_{\mathrm{ext}}$, can be computed 
directly from the definition (\ref{D-dim:length:scale}) as
\begin{align}
	L&=\lim_{T\to0}\left(\fft{\partial S}{\partial T}\right)_{Q_I,X^I_\infty}=\lim_{\rho_0\to0}\left(\fft{\partial T(\rho_0,Q_I,X^I_\infty)}{\partial\rho_0}\right)_{Q_I,X^I_\infty}^{-1}\left(\fft{\partial S(\rho_0,Q_I,X^I_\infty)}{\partial\rho_0}\right)_{Q_I,X^I_\infty}~.\label{5D:STU:length:scales}
\end{align}
This requires some effort because the entropy (\ref{5D:STU:entropy}) and the temperature (\ref{5D:STU:temp}) depend
on $H^I(\rho_0;\rho_0,Q_I,X^I_\infty)$ which in turn are nontrivial functions \eqref{H:def} of $\rho_0$. However, the definition \eqref{5D:STU:length:scales} instructs us 
to take the limit $\rho_0\to 0$ after the differentiations with respect to $\rho_0$ and the complicated looking dependence on $\rho_0$ inside $H^I(\rho_0;\rho_0,Q_I,X^I_\infty)$ is through the combination $\sqrt{\fft14(Q_IX^I_\infty)^2+\rho_0^2}$ which is of second order in $\rho_0^2$. Therefore, from a purely calculational point of view, 
it is only the linear term $\rho=\rho_0$ in $H^I(\rho_0;\rho_0,Q_I,X^I_\infty)$ \eqref{H:def} that contributes and so the final expressions are much simpler than those that appear at intermediate stages. 

The standard nAttractor mechanism automates this simplification, packages it efficiently, and generalizes it to more elaborate settings. The radial function $R$ \eqref{5D:STU:radial} depends on the general functions $H^I(\rho;\rho_0,Q_I,X^I_\infty)$ but the nAttractor formula (\ref{D-dim:length:scale:2}) posits that we 
take the extremal limit $\rho_0=0$ from the outset and immediately recognize $H^I(\rho;0,Q_I,X^I_\infty)$ as a linear function. Only then do we 
compute the length scale
\begin{align}
	L&=\pi S_{\mathrm{ext}}\lim_{\rho\to0}\left(\fft{\partial R(\rho;0,Q_I,X^I_\infty)^3}{\partial\rho}\right)_{Q_I}\nn\\
	&=\fft{\pi^3|Q_1Q_2Q_3|}{16G_5}\left(\fft{1}{|Q_1X^1_\infty|}+\fft{1}{|Q_2X^2_\infty|}+\fft{1}{|Q_3X^3_\infty|}\right)~.\label{5D:STU:L}
\end{align}
This organization of the calculation is much simpler than computation directly from the definition of specific heat 
because it immediately focusses on the extremal solution. 

The microscopic interpretation of the length scale depends on duality frame, i.e. the identification between macroscopic charges and the underlying constituents. 
A canonical assignment for 5D $STU$ black holes is to consider M-theory on a six-torus that is a product of two-tori $T^6 = T^2 \times T^2 \times T^2$ and
interpret charges $Q_i$ in terms of $n_i$ M2-branes wrapping the $i$'th $T^2$ with $i=1, 2, 3$. In this setting the 11D Planck length $\ell_P$ normalizes the 
mass of a single $M_2$-brane as $X^i_\infty \ell_P^{-1}$ where $X^i_\infty = {\cal V}_i/(2\pi\ell_P)^2$ is a dimensionless measure of the volume ${\cal V}_i$ of the 
$i$'th $T^2$. The length scale \eqref{5D:STU:L} becomes
\begin{equation}
L= \pi^2 n_1 n_2 n_3 \ell_P\left( \fft{1}{n_1X^1_\infty} +\fft{1}{n_2X^2_\infty} + \fft{1}{n_3X^3_\infty} \right)~.
\label{5D:STU:micro} 
\end{equation}
In the limit where the constituent mass associated with M2-branes of type 1 is much smaller than that of types 2 and 3, the scale $L$ is set by the modulus 
$X^1_\infty$ of the light M2-brane but it is ``renormalized'' by the number of heavy M2-branes of types 2 and 3. This structure is the hallmark of the ``long'' string scale which plays a central role for black holes in string 
theory \cite{Maldacena:1996ds} 
\footnote{In microscopic applications it is preferable to multiply the overall normalization by $2/\pi$. That was the convention in \cite{Larsen:2018iou}.}. 
The scale \eqref{5D:STU:micro} is a generalization of the long string scale to the generic situation where the M2-branes are on equal footing. The symmetric form was introduced already in \cite{Cvetic:1997uw}.

We can also compute the length scale associated to the horizon values of the scalar fields $X^I$ of the 5D STU near-extremal black hole using the standard nAttractor mechanism. The result is given as
\begin{align}
	L^I&=\fft{2G_5S_{\mathrm{ext}}}{\pi}\lim_{\rho\to0}\left(\fft{\partial X^I(\rho;0,Q_I,X^I_\infty)}{\partial\rho}\right)_{Q_I,X^I_\infty}\nn\\
	&	=\fft{\pi X^I_\infty}{3\sqrt2}\fft{|Q_1Q_2Q_3|^\fft56}{|Q_IX^I_\infty|}\left(\fft{1}{|Q_1X^1_\infty|}+\fft{1}{|Q_2X^2_\infty|}+\fft{1}{|Q_3X^3_\infty|}-\fft{3}{|Q_IX^I_\infty|}\right)~.\label{5D:STU:LI}
\end{align}
Here $X^I$ are constrained to satisfy $X^1X^2X^3=1$ so the length scales $L^I$ are not independent: they satisfy the constraint $Q_IL^I=0$.

\subsection{Black Holes in \texorpdfstring{AdS$_4$}{AdS4}}

In order to study black holes in AdS$_4$ we augment the action \eqref{4D:action} that we already analyzed for asymptotically flat black holes with a potential $\mathcal V[\phi]$ that depends on the scalar fields: 
\begin{align}
	S=&~\fft{1}{16\pi G_4}\int\left[(R^{(4)}-2\mathcal V[\phi])(*\mathbbm{1})-\fft12G_{ij}[\phi](*d\phi^i)\wedge d\phi^j\right.\nn\\&\kern5em\left.-\fft{1}{2}\mu_{IJ}[\phi](*F^I)\wedge F^J-\fft{1}{2}\nu_{IJ}[\phi]F^I\wedge F^J\right]~.\label{Cosmo:action}
\end{align}
The $\mathcal N\geq2$ gauged supergravity coupled to vector and hyper multiplets is a well known example whose bosonic part of the action takes this form. 
We have in mind that the potential allows an AdS$_4$ vacuum $\mathcal V|_{\mathrm{AdS}}=-3g^2$ with scale $\ell_4=g^{-1}$ but results will apply in other settings as well. 
As before, we consider a general spherically symmetric black hole solution to this theory which without loss of generality can be written as
\begin{subequations}
	\begin{align}
	ds^2&=-e^{2\Phi}dt^2+e^{-2\Phi}dr^2+R^2d\Omega^2_2~,\label{Cosmo:ansatz:metric}\\
	P^I\,\mathrm{vol}(S^2)&=F^I|_{S^2}~,\label{Cosmo:ansatz:vector:1}\\
	Q_I\,\mathrm{vol}(S^2)&=\mu_{IJ}(*F^J)|_{S^2}+\nu_{IJ}(F^J)|_{S^2}~,\label{Cosmo:ansatz:vector:2}
	\end{align}\label{Cosmo:ansatz}%
\end{subequations}
where $\Phi$, $R$, and the scalar fields $\phi^i$ are functions of the radial coordinate $r$ and also dependent on the black hole mass $M$ and 
conserved dyonic charges $(P^I,Q_I)$, as well as the asymptotic values of scalar fields $\phi^i_\infty$. 

The ansatz (\ref{Cosmo:ansatz}) automatically satisfies the Bianchi identity and the equation of motion for vector fields $A^I$
\begin{equation}
	dF^I=0\qquad\mbox{and}\qquad d(\mu_{IJ}*F^J+\nu_{IJ}F^J)=0~.
\end{equation}
The scalar equation of motion is 
\begin{equation}
	0=R^{-2}\partial_r(G_{ij}R^2e^{2\Phi}\partial_r\phi^j)-\fft{1}{2}e^{2\Phi}(\partial_iG_{jk})\partial_r\phi^j\partial_r\phi^k
	-\fft{\partial_iV_{\mathrm{eff}}}{2R^4}-2\partial_i\mathcal V~,\label{Cosmo:scalar:eom}
\end{equation}
where the effective potential $V_{\mathrm{eff}}$ is the same as (\ref{4D:eff:potential}) which appeared for asymptotically flat black holes. 
Furthermore, we have the three independent Einstein equations
\begin{subequations}
	\begin{align}
	0&=\partial_r^2(R^2e^{2\Phi})-2+4\mathcal V R^2~,\label{Cosmo:Einstein:1}\\
	0&=\partial_r(e^{2\Phi}R\partial_rR)-1+\fft{1}{4}\fft{V_{\mathrm{eff}}}{R^2}+\mathcal V R^2~,\label{Cosmo:Einstein:2}\\
	0&=\fft{\partial_r^2R}{R}+\fft{1}{4}G_{ij}\partial_r\phi^i\partial_r\phi^j~,\label{Cosmo:Einstein:3}
	\end{align}\label{Cosmo:Einstein}%
\end{subequations}
which are $E_{00}-E_{22}$, $E_{22}$, and $E_{00}+E_{11}$ respectively, where $\{e^0,e^1,\cdots,e^3\}$ denotes the coframe of the metric (\ref{Cosmo:ansatz:metric}). 

In section \ref{D-dim} we analyzed the analogous problem without a potential for the scalar and found that $e^{2\Phi}R^2$ must be quadratic in the 
radial coordinate $r$ introduced in our metric ansatz \eqref{Cosmo:ansatz:metric} .
That is no longer possible because of the term $4\mathcal V R^2$ in the first Einstein equation (\ref{Cosmo:Einstein:1}). We take this into account by writing $e^{2\Phi}$ as 
\begin{equation}
	e^{2\Phi}=\fft{(\rho^2-\rho_0^2)u(\rho)}{R^2}~,\label{new:radial}
\end{equation}
where $u$ is introduced as a function of $\rho$ which is not determined yet. The $\rho$ coordinate is the same as $r$ except that its origin is shifted to remove a possible 
linear term in the quadratic $\rho^2-\rho_0^2$. The asymptotic AdS$_4$ geometry will modify the geometry, but for a near extremal black hole it is reasonable to assume 
that $u$ is smooth and nonvanishing in the near horizon region $\rho\sim\rho_0$. Therefore we can still employ $\rho_0$ in a dual role as both the horizon position and the extremality parameter. We think of $u(\rho)$ as a slowly varying background that allows the cosmological constant to influence quantitative aspects of the near horizon physics but the key qualitative features are unchanged from the asymptotically flat case and remain encoded in the quadratic $\rho^2-\rho_0^2$.

Substituting (\ref{new:radial}) into the metric ansatz (\ref{Cosmo:ansatz:metric}) gives
\begin{equation}
	ds^2=-\fft{(\rho^2-\rho_0^2)u(\rho)}{R^{2}}dt^2+\fft{R^2}{(\rho^2-\rho_0^2)u(\rho)}d\rho^2+R^2d\Omega^2_2~.\label{Cosmo:BH}
\end{equation}
We then rewrite the scalar equation of motion (\ref{Cosmo:scalar:eom}) and the Einstein equations (\ref{Cosmo:Einstein}) using (\ref{new:radial}) as
\begin{subequations}
	\begin{align}
	0=&~\partial_\rho((\rho^2-\rho_0^2)uG_{ij}\partial_\rho\phi^j)-\fft{1}{2}(\rho^2-\rho_0^2)u(\partial_iG_{jk})\partial_\rho\phi^j\partial_\rho\phi^k-\fft{\partial_iV_{\mathrm{eff}}}{2R^{2}}-2R^2\partial_i\mathcal V~,\label{Cosmo:eom:phi}\\
	0=&~2u+2\rho\partial_\rho u+(\rho^2-\rho_0^2)u-2+4\mathcal V R^2~,\label{Cosmo:eom:u}\\
	0=&~\partial_\rho((\rho^2-\rho_0^2)u\partial_\rho\log R)-1+\fft{V_{\mathrm{eff}}}{4R^{2}}+R^2\mathcal V~,\label{Cosmo:eom:R}\\
	0=&~\fft{\partial_\rho^2 R}{R}+\fft{1}{4}G_{ij}\partial_\rho\phi^i\partial_\rho\phi^j~.
	\end{align}\label{Cosmo:eom}%
\end{subequations}
Summarizing so far, we found the general form (\ref{Cosmo:BH}) of spherically symmetric black hole solutions in AdS$_4$, where $u,R,\phi^i$ are determined by (\ref{Cosmo:eom}). In the following we study the equations locally and do not necessarily impose the AdS boundary conditions $R\to\rho$ and $u\to\rho^2 g^2$ 
as $\rho\to\infty$. 

As in section \ref{nAttractor:setting}, we use the horizon position $\rho_0$ instead of the black hole mass $M$ as one of the black hole parameters 
and therefore we write $u,R,\phi^i$ as
\begin{subequations}
\begin{align}
	u&=u(\rho;\rho_0,P^I,Q_I,g,\phi^i_\infty)~,\\
	R&=R(\rho;\rho_0,P^I,Q_I,g,\phi^i_\infty)~,\label{Cosmo:radial}\\
	\phi^i&=\phi^i(\rho;\rho_0,P^I,Q_I,g,\phi^i_\infty)~.
\end{align}
\end{subequations}
We assume that these $u,R,\phi^i$ are smooth with respect to both $\rho$ and $\rho_0$ for $\rho\geq\rho_0$. 

Even though solving the differential equations (\ref{Cosmo:eom}) for $u,R,\phi^i$ remain quite non-trivial, we can derive the algebraic equations that determine their horizon values from (\ref{Cosmo:eom:phi}, \ref{Cosmo:eom:u}, \ref{Cosmo:eom:R}) as
\begin{subequations}
	\begin{align}
	0&=\left[\partial_iV_{\mathrm{eff}}+4R^4\partial_i\mathcal V\right]_{\rho\to\rho_0}~,\\
	0&=\left[u-1+2\mathcal V R^2\right]_{\rho\to\rho_0}~,\label{Cosmo:att:u}\\
	0&=\left[V_{\rm eff}+4R^4\mathcal V-4R^2\right]_{\rho\to\rho_0}~.
	\end{align}
\end{subequations}
These are the attractor equations for the black holes (\ref{Cosmo:BH}).

We can compute the entropy and the temperature of this black hole (\ref{Cosmo:BH}) in terms of the radial function (\ref{Cosmo:radial}) as
\begin{align}
	S(\rho_0,P^I,Q_I,g,\phi^i_\infty)&=\fft{\pi}{G_4} R(\rho_0;\rho_0,P^I,Q_I,g,\phi^i_\infty)^2~,\label{Cosmo:entropy}\\
	T(\rho_0,P^I,Q_I,g,\phi^i_\infty)&=\fft{\rho_0}{2\pi R(\rho_0;\rho_0,P^I,Q_I,g,\phi^i_\infty)^2}~u(\rho_0;\rho_0,P^I,Q_I,g,\phi^i_\infty)~.\label{Cosmo:temp}
\end{align}
Our assumption that $u$ is smooth and nonvanishing in the near-horizon region ensures that the extremal limit $T\to 0$ remain equivalent to $\rho_0\to0$ from (\ref{Cosmo:temp}) also in the presence of a scalar potential.

\subsubsection{Standard nAttractor Mechanism}
As we discussed in subsection \ref{nAttractor:nAttractor}, we encode the near-extremal black hole entropy in the length scale $L$:
\begin{align}
	L&\equiv\lim_{T\to0}\left(\fft{\partial S}{\partial T}\right)_{Q_I}=\fft{\pi}{G_4}\lim_{T\to0}\left(\fft{\partial R(\rho_0;\rho_0,P^I,Q_I,g,\phi^i_\infty)^2}{\partial T}\right)_{P^I,Q_I,g,\phi^i_\infty}~.\label{Cosmo:length:scale}
\end{align}
We seek to compute this length scale using the standard nAttractor mechanism.

First, we trade a derivative with respect to the temperature for a derivative with respect to the horizon position $\rho_0$ and then substitute (\ref{Cosmo:entropy}) and (\ref{Cosmo:temp}) into the resulting expression. The result is given as
\begin{align}
	L&=\lim_{\rho_0\to0}\left(\fft{\partial T(\rho_0,P^I,Q_I,g,\phi^i_\infty)}{\partial\rho_0}\right)^{-1}_{P^I,Q_I,g,\phi^i_\infty}\left(\fft{\partial S(\rho_0,P^I,Q_I,g,\phi^i_\infty)}{\partial\rho_0}\right)_{P^I,Q_I,g,\phi^i_\infty}\nn\\&=\fft{2\pi S_{\mathrm{ext}}}{1-2G_4\mathcal V_{\rm ext}^{\rm hor}S_{\rm ext}/\pi}\lim_{\rho_0\to0}\left(\fft{\partial R(\rho_0;\rho_0,P^I,Q_I,g,\phi^i_\infty)^2}{\partial\rho_0}\right)_{P^I,Q_I,g,\phi^i_\infty}~.\label{Cosmo:length:scale:1}
\end{align}
We took the dependence of the temperature on the function $u$ into account by using the attractor equation (\ref{Cosmo:att:u}) to relate it to $\mathcal V_{\rm ext}^{\rm hor}$, the horizon value of the scalar potential in the extremal limit. This simple maneuver addresses the difficulty posed by the scalar potential. 
We are left with the challenge we have seen repeatedly already, that the $\rho_0$-derivative 
acts on both the 1st (position) and the 2nd argument (black hole parameter) of the radial function $R(\rho_0;\rho_0,P^I,Q_I,g,\phi^i_\infty)$. As in (\ref{D-dim:length:scale:2}) for asymptotically flat black holes, 
it is sufficient to take the $\rho_0$-derivative on the 1st argument (position) into account due to the condition
\begin{equation}
	R(\rho;\rho_0,P^I,Q_I,g,\phi^i_\infty)=R(\rho;0,P^I,Q_I,g,\phi^i_\infty)+\mathcal O(\rho_0^2)~,\label{Cosmo:regularity}
\end{equation}
which follows that (\ref{Cosmo:eom}) depends on $\rho_0$ through $\rho_0^2$ only.

Collecting equations, the standard nAttractor formula for the symmetry breaking scale of a near-extreme black hole in the presence of a scalar potential becomes: 
\begin{align}
	L
	&=\fft{4\pi^2}{G_4}   \fft{R^3_{\rm ext} \partial_\rho R_{\rm ext} }{1-2R^2_{\rm ext}\mathcal V_{\rm ext}^{\rm hor}} ~.\label{Cosmo:length:scale:2}
\end{align}
The $R_{\rm ext}$ refers to the horizon value of the radial function for {\it extremal} black holes, namely $R(0;0,P^I,Q_I,g,\phi^i_\infty)$, and $\partial_\rho R_{\rm ext}$ is a shorthand notation for $\partial_\rho R(\rho;0,P^I,Q_I,g,\phi^i_\infty)$ computed at the extremal horizon $\rho=0$. The potential at the horizon is constant and takes the value $\mathcal V_{\rm ext}^{\rm hor}=-3/l^2$ for AdS$_4$.

In this subsection we have again focused on the length scale $L$ associated with the entropy $S$ of a near-extremal black hole. 
The length scales $L^i$ associated with scalar fields can be obtained analogously by replacing $R^2$ with $\fft{1}{\pi}G_4\phi^i$ 
in (\ref{Cosmo:length:scale}, \ref{Cosmo:length:scale:1}, \ref{Cosmo:length:scale:2}).

\subsubsection{Example: 4D Reissner-Nordstr\"om AdS Black Holes}
The 4D Reissner-Nordstr\"om AdS black hole solution is given by
\begin{subequations}
	\begin{align}
	&ds^2=-e^{2\Phi}dt^2+e^{-2\Phi}dr^2+r^2d\Omega_2^2~,\label{RNAdS:metric}
	\\&e^{2\Phi}=\fft{r^2}{l^2} + 1-\fft{2M}{r}+\fft{Q^2}{4r^{2}}~.
	\end{align}\label{RNAdS:BH}
\end{subequations}
It is a solution to the theory with action (\ref{Cosmo:action}) where
\begin{equation}
	G_{ij}[\phi]=0~(n=0)~,\qquad \mu_{IJ}[\phi]=\delta_{IJ}~(N=1)~,\qquad	\nu_{IJ}[\phi]=0~,\qquad\mathcal V[\phi]=-\fft{3}{l^2}~.
\end{equation}
Thus the scalar potential $\mathcal V[\phi]$ reduces to a cosmological constant. 

To work out the standard nAttractor mechanism for this black hole, we must rewrite the metric (\ref{RNAdS:BH}) in the general form (\ref{Cosmo:BH}) by
introducing a new radial coordinate $\rho=r+c$ where the constant $c$ is arbitrary for now. The $g_{tt}$ component $-e^{2\Phi}$ becomes
\begin{align}
	e^{2\Phi}&=\fft{\fft{1}{l^2}(\rho-c)^4 + (\rho-c)^2-2M(\rho-c)+\fft{1}{4}Q^2}{(\rho-c)^2}\nn\\
	&=\fft{(\rho^2-\rho_0^2)}{(\rho-c)^2}\left(1+\fft{\rho^2-4c\rho+\rho_0^2+6c^2}{l^2}-\fft{2(M+c)+\fft{4c(\rho_0^2+c^2)}{l^2}}{\rho+\rho_0}\right)~,\label{RNAdS:Phi}
\end{align}
where the horizon position $\rho_0$ is defined as the largest root of $e^{2\Phi}$, namely
\begin{equation}
	\fft{1}{l^2}(\rho_0-c)^4 +  (\rho_0-c)^2-2M(\rho_0-c)+\fft{Q^2}{4}=0~.\label{RNAdS:horizon}
\end{equation}
Comparing (\ref{RNAdS:metric}) and (\ref{RNAdS:Phi}) with the general expressions (\ref{Cosmo:BH}) and (\ref{new:radial}), respectively, we have
\begin{subequations}
\begin{align}
	R(\rho;\rho_0,Q,l)&=\rho-c~,\label{RNAdS:radial}\\
	u(\rho;\rho_0,Q,l)&=1+\fft{\rho^2-4c\rho+\rho_0^2+6c^2}{l^2}-\fft{2(M+c)+\fft{4c(\rho_0^2+c^2)}{l^2}}{\rho+\rho_0}~,\label{RNAdS:u}
\end{align}\label{RNAdS:general}%
\end{subequations}
where $c$ will be computed shortly in terms of $\rho_0,Q,l$ by demanding the smoothness of the function $u(\rho;\rho_0,Q,l)$ below and $M$ is also a function of $\rho_0,Q,l$ which can be obtained explicitly by substituting the expression $c(\rho_0,Q,l)$ into (\ref{RNAdS:horizon}). In summary, we rewrote the RN AdS black hole metric (\ref{RNAdS:BH}) in the general form (\ref{Cosmo:BH}) with $R,u$ given as (\ref{RNAdS:general}).

We now determine the constant $c$. Since $u(\rho;\rho_0,Q,l)$ must be a smooth function of both $\rho$ and $\rho_0$ for $\rho\geq\rho_0$ even at extremality $\rho=\rho_0\to0$, the last term of (\ref{RNAdS:u}) vanishes. Combining this condition with the constraint (\ref{RNAdS:horizon}), we can determine the constant $c$ in terms of $\rho_0,Q,l$ as
\begin{align}
	&\begin{cases}
	0=2(M+c)+\fft{4c(\rho_0^2+c^2)}{l^2}~,\\
	0=\fft{1}{l^2}(\rho_0-c)^4 + (\rho_0-c)^2-2M(\rho_0-c)+\fft{1}{4}Q^2~,
	\end{cases}\nn\\
	&\Rightarrow\quad
	c(\rho_0,Q,l)= - \fft{l}{\sqrt6}\left(\sqrt{\left(1+\fft{4\rho_0^2}{l^2}\right)^2+\fft{3Q^2}{l^2}} -\left(1-\fft{2\rho_0^2}{l^2}\right) \right)^\fft12~.\label{shift}
\end{align}
We chose the negative root for $c$ so the radial function $R_{\rm ext} = R(0;0,Q,l)=-c(0,Q,l)$ is positive at the extremal horizon. Our explicit result for $c$ 
verifies that in this example the radial function $R(\rho;\rho_0,Q,l)$ (\ref{RNAdS:radial}) is smooth with respect to both $\rho$ and $\rho_0$ for $\rho\geq\rho_0$, as we expect. Moreover, 
the dependence on a small $\rho_0$ is quadratic, as we demand in (\ref{Cosmo:regularity}). 

We have exhibited all details in this example, to illustrate our reasoning and justify the various steps. However, for a practical computation the essential ingredients are
that the extremal radial function $R(\rho;0,Q,l)$ depends linearly on $\rho$ and that its horizon value is given as 
$$
R_{\rm ext} = \fft{l}{\sqrt6}\left(\sqrt{1+\fft{3Q^2}{l^2}} - 1\right)^\fft12~.
$$
The nAttractor equation (\ref{Cosmo:length:scale:2}) for gauged supergravity then immediately gives 
\begin{align}
	L=\fft{4\pi^2}{G_4} \fft{R^3_{\rm ext}\partial_\rho R_{\rm ext}}{1+6R^2_{\rm ext}/l^2} 	=\fft{4\pi^2l^3}{G_4}  \fft{\left(\sqrt{1+3Q^2/l^2} - 1 \right)^\fft32}{6^{\frac{3}{2}}\sqrt{1+3Q^2/l^2}}~.
\end{align}
%

\subsection{Rotating Black Holes}
In this subsection we develop a nAttractor mechanism for spinning black holes. Rotating attractors are much more complicated than spherical symmetric ones already in the extremal limit \cite{Astefanesei:2006dd}. We therefore limit the scope slightly and focus on 4D from the outset and we only analyze the entropy nAttractor, leaving scalar fields for later study. However, 
we consider again the generic 4D action (\ref{4D:action}) that couples gravity to electromagnetic fields and scalars. 

We need to address the general stationary, axisymmetric (Weyl-Lewis-Papapetrou) ansatz
\begin{equation}
	ds^2=-e^{2U(r,z)}(dt+A(r,z)d\phi)^2+e^{-2U(r,z)}r^2d\phi^2+e^{2V(r,z)}(dr^2+dz^2)~.
\end{equation}
Here $U$, $V$, $A$ are not just functions of coordinates $\rho,z$, they also depend on the black hole data: the mass $M$, conserved dyonic charges $(P^I,Q_I)$, and the angular momentum $J$. Those parameters were omitted in their arguments for notational convenience. Under the change of variables $(r,z)=(\sqrt{\rho^2-\rho_0^2}\sin\theta,\rho\cos\theta)$, the general metric can be rewritten as
\begin{equation}
	ds^2=-\fft{H(\rho,\theta)}{W(\rho,\theta)}\left(dt+\mathcal A(\rho,\theta)d\phi\right)^2+W(\rho,\theta)\left(\fft{d\rho^2}{\rho^2-\rho_0^2}+d\theta^2+\fft{\rho^2-\rho_0^2}{H(\rho,\theta)}\sin^2\theta d\phi^2\right)~,\label{Rot:BH}
\end{equation}
where we have introduced
\begin{align}
	W(\rho,\theta)&= e^{2V(r,z)}(\rho^2-\rho_0^2\cos^2\theta)~,\\
	H(\rho,\theta)&= e^{2U(r,z)+2V(r,z)}(\rho^2-\rho_0^2\cos^2\theta)~,\\
	\mathcal A(\rho,\theta)&= A(r,z)~.
\end{align}
The new coordinates were chosen so there is an event horizon at $\rho=\rho_0$, as in previous examples. 

The fact that we consider a black hole restricts the $\theta$-dependence on the horizon $\rho=\rho_0$ so that
\begin{subequations}
	\begin{align}
	\mathcal A(\rho_0,\theta)&=-\fft{1}{\Omega(\rho_0)}~,\label{Rot:assumption:A}\\
	H(\rho_0,\theta)&=-h(\rho_0)\sin^2\theta\leq 0~,\label{Rot:assumption:H}
	\end{align}\label{Rot:assumption:1}%
\end{subequations}
where $\Omega(\rho_0)$ and $h(\rho_0)$ are functions of the horizon position $\rho_0$ but independent of the angular coordinate $\theta$. The first condition (\ref{Rot:assumption:A}) is required for the event horizon $\rho=\rho_0$ to coincide with the Killing horizon of the Killing vector $\chi=\partial_t+\Omega(\rho_0)\partial_\phi$. The second condition (\ref{Rot:assumption:H}) is necessary for a well-defined temperature which is positive and uniform ($\theta$-independent) over the event horizon as
\begin{equation}
	T=\fft{\kappa}{2\pi}=\fft{\sqrt{-\fft{1}{2}\nabla_\mu\chi_\nu\nabla^\mu\chi^\nu}}{2\pi}=\fft{1}{2\pi}\sqrt{-\fft{\rho_0^2\Omega(\rho_0)^2\sin^2\theta}{H(\rho_0,\theta)}}=\fft{\rho_0|\Omega(\rho_0)|}{2\pi\sqrt{h(\rho_0)}}~.
\end{equation}
The inequality $H\leq 0$ indicates that the event horizon is situated inside an ergoregion. 

According to the organization of previous subsections, at this point we should extend functions like (\ref{Rot:assumption:A}-\ref{Rot:assumption:H}) away from the horizon by studying the Bianchi identity and the equation of motion for vector fields $A^I$, the equation of motion for scalar fields $\phi^i$, and the Einstein equations for 
the general metric ansatz. However, such an analysis is much more difficult without spherical symmetry and we shall not attempt it here. Instead we proceed by making reasonable assumptions needed to formulate a nAttractor mechanism for rotating black holes and then subject our proposal to nontrivial tests, but leave a detailed justification for the future. 

In the previous subsections we computed black hole entropy and temperature in terms of a radial function $R$ that encoded all of the black hole geometry. However, 
there is no natural ``radial function'' in the metric (\ref{Rot:BH}) without spherical symmetry. Instead, we construct a radial function $R$ as
\begin{equation}
	4\pi R(\rho;\rho_0,P^I,Q_I,J,\phi^i_\infty)^2\equiv\int d\theta d\phi\sqrt{-H(\rho,\theta)\mathcal A(\rho,\theta)^2\left(1-\fft{W(\rho,\theta)^2(\rho^2-\rho_0^2)\sin^2\theta}{H(\rho,\theta)^2\mathcal A(\rho,\theta)^2}\right)}~,\label{Rot:radial}
\end{equation}
which is the area of a 2-dimensional surface defined by $t={\rm constant}$ and $\rho={\rm constant}$. We have again eliminated 
the black hole mass $M$ in favor of the horizon position $\rho_0$. 

We can compute the entropy and the temperature of the black hole (\ref{Rot:BH}) in terms of the radial function (\ref{Rot:radial}) as 
\begin{align}
	S(\rho_0,P^I,Q_I,J,\phi^i_\infty)&=\fft{\pi}{G_4} R(\rho_0;\rho_0,P^I,Q_I,J,\phi^i_\infty)^2~,\label{Rot:entropy}\\
	T(\rho_0,P^I,Q_I,J,\phi^i_\infty)&=\fft{\rho_0}{2\pi R(\rho_0;\rho_0,P^I,Q_I,J,\phi^i_\infty)^2}~.\label{Rot:temp}
\end{align}
Regular black holes have strictly positive entropy so $R(\rho_0;\rho_0,P^I,Q_I,J,\phi^i_\infty)>0$ from (\ref{Rot:entropy}). The extremal limit $T\to0$ is therefore equivalent to $\rho_0\to0$ from (\ref{Rot:temp}). We shall assume that the $R$ defined above is smooth with respect to both $\rho$ and $\rho_0$ for $\rho\geq\rho_0$. 
Then the entropy of a near-extremal black hole will smoothly approach the extremal value $S_{\mathrm{ext}}=S(0,P^I,Q_I,J,\phi^i_\infty)$ as the temperature is lowered.

The complicated looking integral in the definition (\ref{Rot:radial}) of the radial function simplifies significantly at the horizon $\rho=\rho_0$. There the 
entire $\theta$-dependence is an overall $\sin\theta$, as in flat geometry, and the integral over angles gives $4\pi$, as for an elementary unit sphere. 
Therefore the entropy (\ref{Rot:entropy}) and the temperature (\ref{Rot:temp}) are easily computed in explicit examples, as we illustrate in 
subsection \ref{CY:BH}. Our working hypothesis is that similar simplifications remain also infinitesimally away from the horizon. Since the horizon sphere is effectively round in our coordinates it is reasonable to expect that it remains so after infinitesimal radial motion away from the horizon. 

\subsubsection{Standard nAttractor Mechanism}
The standard nAttractor mechanism for a rotating black hole can be established exactly as in subsections \ref{nAttractor:nAttractor} and \ref{D-dim:nAttractor}. To be specific, we define the length scale quantifying the entropy of a rotating near-extremal black hole as
\begin{equation}
	L\equiv\lim_{T\to0}\left(\fft{\partial S}{\partial T}\right)_{P^I,Q_I,J,\phi^i_\infty}=\fft{\pi}{G_4}\lim_{\rho_0\to0}\left(\fft{\partial R(\rho_0;\rho_0,P^I,Q_I,J,\phi^i_\infty)^2}{\partial\rho_0}\right)_{P^I,Q_I,J,\phi^i_\infty}~.\label{Rot:length:scale}
\end{equation}
Then the standard nAttractor mechanism states that this length scale can be computed directly in the
extremal geometry
\begin{align}
	L&=2\pi S_{\mathrm{ext}}\lim_{\rho\to0}\left(\fft{\partial R(\rho;0,P^I,Q_I,J,\phi^i_\infty)^2}{\partial\rho}\right)_{P^I,Q_I,J,\phi^i_\infty}\nn\\
	&= - \fft{\pi}{4G_4}\int d\theta d\phi\,\lim_{\rho\to0}\left(\fft{\partial_\rho H(\rho,\theta)+2h(0)\Omega(0)\sin^2\theta\,\partial_\rho\mathcal A(\rho,\theta)}{\Omega(0)^2\sin\theta}\right)_{\rho_0=0}~.\label{Rot:length:scale:2}
\end{align}
To reach the final line we substituted (\ref{Rot:assumption:1}, \ref{Rot:radial}) into the 1st line and assumed that $W(\rho,\theta)$ is smooth near $\rho=0$. 
This is justified if 
\begin{equation}
	R(\rho;\rho_0,P^I,Q_I,J,\phi^i_\infty)=R(\rho;0,P^I,Q_I,J,\phi^i_\infty)+\mathcal O(\rho_0^2)~.\label{Rot:regularity}
\end{equation}
However, since we did not investigate the equations of motion as we did in the previous sections we have just {\it formulated} the standard nAttractor mechanism for 
general rotating black holes; we have not {\it established} it. In the following subsection, we consider an explicit example where we can show that the standard nAttractor 
mechanism is satisfied. Additionally, it was already noticed that analogous simplifications hold for near-BPS rotating black holes in AdS$_5$ \cite{Larsen:2019oll}.

The nAttractor formula \eqref{Rot:length:scale:2} simplifies further in some important situations. 
It often happens that $H$ departs only quadratically in the coordinate $\rho$ from its horizon value \eqref{Rot:assumption:H}. Then the derivative
$\partial_\rho H=0$. Furthermore, the rotational potential ${\cal A}$ may be corrected from its horizon value \eqref{Rot:assumption:A} by a term that is linear in $\rho$ but independent of the polar angle angle $\theta$, thus corresponding to a constant ``rotational'' field strength. Under these assumptions the nAttractor formula becomes
\begin{equation}
L =  - \fft{2\pi^2}{G_4} \frac{h(0) \partial_\rho {\cal A}}{ \Omega}~.
\label{eqn:Linertia}
\end{equation}
The field strength in the numerator is proportional to the angular momentum $J$ so the entire expression can be identified with 
the {\it moment of inertia} of the extremal black hole. 

\subsubsection{Example: Cvetic-Youm 4-Charge Rotating Black Hole}\label{CY:BH}
Here we consider the rotating, 4-charge Cvetic-Youm solution \cite{Cvetic:1996kv}, following the convention of \cite{Chow:2013tia}. The black hole metric is given as 
\begin{align}
	ds^2=&-\fft{r^2-2mr+a^2\cos^2\theta}{W}\left(dt+\fft{Ua\sin^2\theta}{r^2-2mr+a^2\cos^2\theta}d\phi\right)^2\nn\\&+W\left(\fft{dr^2}{r^2-2mr+a^2}+d\theta^2+\fft{r^2-2mr+a^2}{r^2-2mr+a^2\cos^2\theta}\sin^2\theta d\phi^2\right)~.
\end{align}
The physical black hole variables $M,Q_I,J$ with $I=1,2,3,4$ are parametrized by $m,\delta_I,a$ as $(s_I\equiv\sinh\delta_I,~c_I\equiv\cosh\delta_I)$
\begin{align}
	M=~m\left(1+\fft12\Sigma_{I=1}^4s_I^2\right)~,\quad Q_I=2ms_Ic_I~,\quad J=ma\left({\scriptstyle\prod_{I=1}^4}c_I-{\scriptstyle\prod_{I=1}^4}s_I\right)~.\label{CY:parameters}
\end{align}
The metric functions $U,W$ are given as 
\begin{subequations}
	\begin{align}
	U=&~2(J/a)r+4m^2{\scriptstyle\prod_{I=1}^4}s_I~,\label{eqn:Udef} \\
	W^2=&~(r^2-2mr+a^2\cos^2\theta)^2+U^2+4M(r^2-2mr+a^2\cos^2\theta)(r-a\cos\theta)\nn\\&+4m^2(r^2-2mr+a^2\cos^2\theta)\left(1+\Sigma_Is_I^2+\Sigma_{I<J}s_{I}^2s_{J}^2-(J/ma)^2\right)~.
	\end{align}\label{CY:BH:1}%
\end{subequations}
This black hole solution takes the general form (\ref{Rot:BH}) if we rewrite it in terms of a new radial coordinate $\rho=r-m$ and use the following identifications,
\begin{align}
	\rho_0^2=m^2-a^2~,\quad H(\rho,\theta)=\rho^2-\rho_0^2-a^2\sin^2\theta~,\quad\mathcal A(\rho,\theta)=\fft{Ua\sin^2\theta}{\rho^2-\rho_0^2-a^2\sin^2\theta}~.\label{CY:BH:2}
\end{align}
Recall that rotating black holes are not inherently equipped with a radial function $R$, but we introduced a proxy in (\ref{Rot:radial}).  
Upon insertion of (\ref{CY:BH:1}, \ref{CY:BH:2}) it becomes
\begin{subequations}
\begin{align}
	R(\rho;\rho_0,Q_I,J,\phi^i_\infty)^2=&~\fft{1}{4\pi}\int d\theta d\phi\,\mathcal R(\rho;\rho_0,\theta,Q_I,J,\phi^i_\infty)^2\sin\theta~,\label{Rot:Cvetic:radial}\\
	\mathcal R(\rho;\rho_0,\theta,Q_I,J,\phi^i_\infty)^4=&~\fft32\rho^2a^2+\fft12a^4+\fft12a^2(\rho^2-\rho_0^2)\cos2\theta+\prod_I(\rho+\tilde Q_I)+\rho a^2\sum_I\tilde Q_I\nn\\&-\fft{a^2}{2(\rho_0^2+a^2)}\left(\prod_I\tilde Q_I-\prod_I |Q_I|\right)+\fft{a^2}{2}\sum_{I<J}\tilde Q_I\tilde Q_J~.
\end{align}\label{CY:radial}%
\end{subequations}
Here we have defined $\tilde Q_I\equiv\sqrt{\rho_0^2+a^2+Q_I^2}$ for notational convenience.

Already at this point, you can expect that the standard nAttractor mechanism works in this example since the radial function (\ref{CY:radial}) is smooth with respect to both $\rho$ and $\rho_0$ for $\rho\geq\rho_0$ and also satisfies the condition (\ref{Rot:regularity}). Here we show explicitly how it works though. First, we can compute the entropy and the temperature of this black hole in terms of the radial function (\ref{CY:radial}) using (\ref{Rot:entropy}, \ref{Rot:temp}) as
\begin{align}
	S(\rho_0,Q_I,J,\phi^i_\infty)&=\fft{\pi}{G_4}R(\rho_0;\rho_0,Q_I,J,\phi^i_\infty)^2~,\label{CY:entropy}\\
	T(\rho_0,Q_I,J,\phi^i_\infty)&=\fft{\rho_0}{2\pi R(\rho_0;\rho_0,Q_I,J,\phi^i_\infty)^2}~.\label{CY:temp}
\end{align}
The entropy (\ref{CY:entropy}) is strictly positive provided that $a\neq0$ from (\ref{CY:radial}). Under the same condition, the extremal limit $T\to0$ is equivalent to $\rho_0\to0$: the horizon parameter $\rho_0$ is an extremality parameter. The entropy of an extremal black hole, which we abbreviate as $S_{\mathrm{ext}}$, is therefore given as $S_{\mathrm{ext}}=S(0,Q_I,J,\phi^i_\infty)$.

In principle, the length scale associated with the near-extremal entropy of the Cvetic-Youm 4-charge rotating black hole can be computed directly from 
the definition (\ref{Rot:length:scale}) as
\begin{equation}
	L\equiv\lim_{T\to0}\left(\fft{\partial S}{\partial T}\right)_{Q_I,J,\phi^i_\infty}=\lim_{\rho_0\to0}\left(\fft{\partial T(\rho_0,Q_I,J,\phi^i_\infty)}{\partial\rho_0}\right)^{-1}_{Q_I,J,\phi^i_\infty}\left(\fft{\partial S(\rho_0,Q_I,J,\phi^i_\infty)}{\partial\rho_0}\right)_{Q_I,J,\phi^i_\infty}~.
\end{equation}
However, this requires a heavy calculation because the entropy and the temperature given in (\ref{CY:entropy}, \ref{CY:temp}) are written in terms of the radial function (\ref{CY:radial}), whose partial derivative with respect to $\rho_0$ with fixed $Q_I,J,\phi^i_\infty$ is highly complicated. The main obstacle is that
(\ref{CY:radial}) is written in terms of $a$, instead of $J$, which is a complicated function of $\rho_0$, $Q_I$, and $J$ through (\ref{CY:parameters}), namely
\begin{equation}
	a=a(\rho_0,Q_I,J)~~\mbox{satisfying}~~ Q_I=2\sqrt{\rho_0^2+a^2}s_Ic_I~~\mbox{and}~~J=a\sqrt{\rho_0^2+a^2}(c_{1234}-s_{1234}).
\end{equation}
Computing the partial derivative of $R(\rho_0;\rho_0,Q_I,J,\phi^i_\infty)$ with respect to $\rho_0$ with fixed $Q_I,J,\phi^i_\infty$ is therefore much more involved than computing the similar partial derivative with fixed $Q_I,a,\phi^i_\infty$.

In contrast, the standard nAttractor mechanism gives the simple formula (\ref{Rot:length:scale:2}) for the effective length scale. 
Moreover, the present example satisfies the additional assumption leading to the ``moment of inertia'' formula for the nAttractor scale \eqref{eqn:Linertia}. 
We therefore easily find
\begin{align}
	L=  \fft{4\pi^2}{G_4} \fft{J}{\Omega} = \fft{8\pi^2 a^3}{G_4} \left( \prod_{I=1}^4\cosh^2\delta_I - \prod_{I=1}^4\sinh^2\delta_I\right)~,
	\label{CY:length:scale}
\end{align}
where we identified $J= - \fft12 h(0)\partial_\rho{\cal A}$. 
The evaluation of the nAttractor formula is clearly much simpler than the calculation starting from the definition of specific heat. The main ``practical'' advantage is that the parametric form of physical variables makes it quite laborious to maintain $Q_I,J,\phi^i_\infty$ fixed when lowering the temperature.  But there is no analogous challenge when differentiating with respect to the radial coordinate in the extremal geometry. 

The length scale (\ref{CY:length:scale}) applies to the general rotating black hole with four independent charges. It is not possible to invert the parametric form and write the result in terms of physical charges $Q_I, J$ alone. As a check on the computations, the result agrees with the one computed in \cite{Larsen:2018iou} for nonrotating black holes.

\section{The Strong nAttractor Mechanism in 5D \texorpdfstring{$\mathcal N=2$}{N=2} Ungauged Supergravity}
\label{strong}

In this section we develop the {\it strong} nAttractor mechanism.
It gives the scales of the near-extremal black hole entirely in terms of the data of the extremal attractor mechanism and the asymptotic scalars, as in (\ref{N=2:4D:STEP2}).

Starting from the standard nAttractor, we need to recast the geometric derivative that acts along the flow away from the extremal attractor as a gradient in charge space contracted with an appropriate ``normal'' vector. This trade was implemented explicitly for BPS black holes in 4D $\mathcal N=2$ ungauged supergravity \cite{Larsen:2018iou}, where the BPS black holes can be written entirely in terms of linear functions of the form $H_I=q_I^\infty r+Q_I$ and $\tilde H^I=p^I_\infty r+P^I$, in which case we have
\begin{equation}
    \fft\partial{\partial r}\to\sum_I q_I^\infty\fft\partial{\partial Q_I}+p^I_\infty\fft\partial{\partial P^I}~.
    \label{eq:rbyQ}
\end{equation}
In addition, the required relation between the quantities $(p^I_\infty,q_I^\infty)$ and the asymptotic values of the scalars $z^i_\infty$ was developed. 

We thus expect that the strong nAttractor mechanism should hold at least whenever the corresponding extremal black holes can be written in terms of linear functions of an appropriate radial coordinate, and the parameters in the linear functions can be related to asymptotic charges and scalar VEVs.  However, since the structure of extremal black hole solutions and the moduli space of scalars are both model dependent, it is not yet clear what degree of universal the strong nAttractor mechanism enjoys.  As a step towards a more general framework, in this section we extend the strong nAttractor mechanism for 4D near-BPS black holes \cite{Larsen:2018iou} to the case of near-extremal black hole in 5D $\mathcal N=2$ ungauged supergravity, both near-BPS and near non-BPS. 

\subsection{The Attractor Mechanism for Extremal Black Holes}\label{5D:N=2}

We focus on static, spherically symmetric, electrically charged black holes in 5D $\mathcal N=2$ ungauged supergravity coupled to $N$ vector multiplets. We will refer to such black holes simply as ``black holes'' in this section. The bosonic action and general features of the theory were discussed in subsection~\ref{sec:D5N2sugra}. There we also demonstrated that the standard nAttractor mechanism gives the scales of general {\it near-extremal} black holes in this theory as radial derivatives acting on functions that specify the {\it extremal} black hole solutions. The strong nAttractor mechanism that we develop in this section therefore exclusively involves analysis of the corresponding extremal black holes.

Extremal black holes in 5D $\mathcal N=2$ ungauged supergravity are obtained by solving the coupled differential equations (\ref{5D:sugra:eom}) with $\rho_0=0$. Recall that the effective scalar potential $V_{\rm eff}$ and the cubic prepotential $\mathcal V$ in these equations of motion are given as
\begin{align}
	V_{\mathrm{eff}}&=(G^{-1})^{IJ}Q_IQ_J~,\label{eq:Veff}\\
	\mathcal V&=\fft16C_{IJK}X^IX^JX^K~,\label{eqn:prepot}
\end{align}
respectively, where the $N+1$ real scalar fields $X^I$ are constrained by $\mathcal V=1$.

At the extremal horizon $\rho=\rho_0=0$ the scalar equation of motion (\ref{5D:sugra:eom:scalar}) reduces to 
\begin{align}
	\fft\partial{\partial {X^I}}\left(V_{\mathrm{eff}}+\lambda\mathcal V\right)=0~,\label{5D:sugra:att:eqn}
\end{align}
where $\lambda$ is the Lagrange multiplier that imposes the constraint $\mathcal V=1$ after differentiation. Then the first Einstein equation (\ref{5D:sugra:eom:scalar}) yields $R^4=V_{\rm eff}/12$ at the extremal horizon and hence determines the horizon value of the radial function. These steps constitute the ordinary attractor mechanism, which applies to extremal black holes without requiring them to be BPS. The practical challenge is that the effective scalar potential (\ref{eq:Veff}) depends on the inverse of the kinetic function $G_{IJ}$ given in (\ref{5D:sugra:metric}) which in general does not have a useful form.

In order to simplify the attractor equation (\ref{5D:sugra:att:eqn}), we restrict to the case where the scalar manifold is homogeneous and symmetric. In this situation, the structure constants $C_{IJK}$ of the prepotential \eqref{eqn:prepot} satisfy (see e.g. \cite{Gutowski:2004yv})
\begin{equation}
	C^{IJK}C_{J(LM}C_{NO)K}=\fft43\delta^I{}_{(L}C_{MNO)}~,\label{CIJK:sym}
\end{equation}
where we use $\delta^{IJ}$ as a raising operator. Using this constraint, we can derive the following useful expressions for $X^I$ and $(G^{-1})^{IJ}$
\begin{align}
	X^I&=\fft92C^{IJK}X_JX_K~,\label{XI:sym}\\
	(G^{-1})^{IJ}&=X^IX^J-3C^{IJK}X_K~,\label{5D:sugra:metric:inverse}
\end{align}
in terms of $X_I$ defined in (\ref{5D:sugra:lower}). Simplifying (\ref{eq:Veff}) using (\ref{5D:sugra:metric:inverse}) and differentiating with respect to $X^I$ then gives the explicit form of the attractor equation (\ref{5D:sugra:att:eqn})
\begin{equation}
	2X^JQ_JQ_I-C_{IJK}X^JC^{KLM}Q_LQ_M+3\lambda X_I=0~.\label{5D:sugra:att:eqn:sym}
\end{equation}
The Lagrange multiplier $\lambda$ may be eliminated by contracting this equation with $X^I$ and using the constraint $X_IX^I=1$.  The resulting expression is non-linear in the horizon value of the scalars $X^I$, and hence somewhat involved to work with.  Nevertheless, it illustrates the extremal attractor mechanism where 
some and possibly all of the scalars are fixed at the horizon in terms of the charges $Q_I$.

The most general extremal black holes in 5D $\mathcal N=2$ ungauged supergravity with a homogeneous symmetric scalar manifold, whose scalars flow to solutions of the attractor equations (\ref{5D:sugra:att:eqn:sym}) at the horizon, are not yet known explicitly so in the following we consider two distinct subcases. The first is for the BPS (extremal) black holes for arbitrary homogeneous symmetric scalar manifolds and the second is for all extremal (BPS and non-BPS) black holes for the specific homogeneous symmetric scalar manifold corresponding to the five-dimensional $ST(N)$ model. 

\subsection{BPS Black Holes}
\label{5D:N=2:STEP2:BPS}

While the extremization equation (\ref{5D:sugra:att:eqn:sym}) is complicated in general, for supersymmetric black holes it is solved exactly when the BPS attractor equation is satisfied \cite{Sabra:1997yd}
\begin{equation}
    X_I=\fft{Q_I}Z\qquad\mbox{where}\qquad Z=Q_JX^J~.
    \label{eq:BPSattractor}
\end{equation}
The proportionality constant $Z$ can be computed by inserting this into the prepotential constraint $\mathcal V=1$. It is given by 
\begin{equation}
    Z^3=\fft92C^{IJK}Q_IQ_JQ_K~.
    \label{eq:Zcc}
\end{equation}
(Here we are still assuming a homogeneous symmetric scalar manifold.)

In fact, the full black hole solution was obtained in \cite{Sabra:1997yd} by solving the first order BPS equations.  The result can be expressed as
\begin{subequations}
	\begin{align}
	ds^2&=-\fft{\rho^2}{R^4}dt^2+\fft{R^2}{4\rho^2}d\rho^2+R^2d\Omega_3^2~,\label{5D:sugra:metric:BPS}\\
	*F^I&=(G^{-1})^{IJ}Q_J\,\mathrm{vol}(S^3)~,\label{5D:sugra:vector:BPS}\\
	X_I&=\fft{H_I}{6R^2}~,\label{5D:sugra:scalar:BPS}
	\end{align}\label{5D:sugra:BPS}
\end{subequations}
where the radial function $R$ is given by
\begin{equation}
	R^6=\fft{1}{48}C^{IJK}H_IH_JH_K~.\label{5D:sugra:radial:BPS}
\end{equation}
A key feature of this BPS solution is that it is built from a set of $N+1$ linear functions
\begin{equation}
    H_I=q_I^\infty\rho+Q_I~.\label{eq:linfun}
\end{equation}
At the horizon $\rho =0$ we have $H_I=Q_I$ so  
comparison of (\ref{5D:sugra:radial:BPS}) and (\ref{eq:Zcc}) 
determines $R^2=Z/6$ at the horizon, and the scalars in (\ref{5D:sugra:scalar:BPS}) flow to their attractor values at the horizon, as they should.  This solution preserves exactly half of the $\mathcal N=2$ supersymmetries; a solution preserving the other half may be obtained by taking instead
\begin{equation}
X_I=-\fft{H_I}{6R^2}~,\qquad R^6=-\fft{1}{48}C^{IJK}H_IH_JH_K~,
\end{equation}
which gives $R^2|_{\mathrm{hor}}=-Z/6$.

The linear functions \eqref{eq:linfun} are essentially the ``harmonic'' functions that underpin many BPS solutions in supergravity (and that is the origin of the notation ``$H_I$''). They differ by an overall factor $\rho=r^2$ and $Q_I/r^2$ is indeed harmonic in four flat spatial dimensions with metric $ds^2=dr^2 + r^2d^2\Omega_3$. It is common to pick ``canonical'' integration constants so the harmonic functions are $1 + Q_I/r^2$, but in our context it is important that we keep general moduli. 

At this point it is clear how the strong nAttractor mechanism works for the near-extremal black holes whose extremal limit is the BPS solution given in (\ref{5D:sugra:BPS}).  Since the radial function (\ref{5D:sugra:radial:BPS}) and the scalar fields (\ref{5D:sugra:scalar:BPS}) are written exclusively in terms of linear functions $H_I$, we can replace a radial derivative on $R$ and $X^I$ with a gradient with respect to conserved charges $Q_I$ contracted with a ``normal vector'' $q_I^\infty$. Therefore, the strong nAttractor mechanism recasts the overall scale $L$ given in \eqref{D-dim:length:scale:2} as
\begin{align}
	L&\overset{\scriptscriptstyle\mathrm{standard}}{=}\pi S_{\mathrm{ext}}\lim_{\rho\to0}\left(\fft{\partial R(\rho;0,Q_I,X^I_\infty)^3}{\partial\rho}\right)_{Q_I,X^I_\infty}\nn\\&~\overset{\scriptscriptstyle\mathrm{strong}}{=}\pi S_{\mathrm{ext}}\sum_{I=0}^Nq_I^\infty\fft{\partial R(0;0,Q_I,X^I_\infty)^3}{\partial Q_I}~,\label{5D:sugra:STEP2L:BPS}
\end{align}
and the scales $L^I$ characterizing the approach of the scalar fields to their attractor values at the horizon become
\begin{align}
	L^I&\overset{\scriptscriptstyle \mathrm{standard}}{=}\fft{2G_5S_{\mathrm{ext}}}{\pi}\lim_{\rho\to0}\left(\fft{\partial X^I(\rho;0,Q_J,X^J_\infty)}{\partial\rho}\right)_{Q_J,X^J_\infty}\nn\\&~\overset{\scriptscriptstyle \mathrm{strong}}{=}\fft{2G_5S_{\mathrm{ext}}}{\pi}\sum_{J=0}^Nq_J^\infty\fft{\partial X^I(0;0,Q_J,X^J_\infty)}{\partial Q_J}~.\label{5D:sugra:STEP2LI:BPS}
\end{align}
Since the $X^I$'s are constrained scalars, the corresponding length scales $L^I$ are not independent, but satisfy the constraint $Q_IL^I=0$.

Of course, we still have to determine the ``normal'' vector $q_I^\infty$ in terms of the asymptotic scalars $X^I_\infty$.  As in the 4D $\mathcal N=2$ case considered in \cite{Larsen:2018iou}, the linear nature of $H_I$ along with homogeneity ensure that the asymptotic values of the scalars follow from the BPS attractor equation (\ref{eq:BPSattractor}) with the replacement $Q_I\to q_I^\infty$.  In particular, we have
\begin{equation}
	X_I^\infty=\fft{q_I^\infty}{Z_\infty}\qquad\mbox{where}\qquad
	(Z_\infty)^3=\fft92C^{IJK}q_I^\infty q_J^\infty q_K^\infty~.\label{Xinf:qinf}
\end{equation}
These equations do not yield a unique inverse relation as there is a scaling symmetry taking $q_I^\infty\to \lambda q_I^\infty$.  
The lack of an inverse is in fact expected, as there are $N+1$ free parameters $q_I^\infty$ while there are only $N$ independent 
scalars because of the prepotential constraint. This overall scale symmetry can be removed by imposing asymptotic flatness $R^2\to\rho$ as $\rho\to\infty$. This is equivalent to taking $Z_\infty=6$, which determines $q_I^\infty$ in terms of $X^I_\infty$ simply as
\begin{equation}
    q_I^\infty=6X_I^\infty=C_{IJK}X^J_\infty X^K_\infty~.
    \label{eq:qinfty}
\end{equation}

Finally, substituting this result into (\ref{5D:sugra:STEP2L:BPS}, \ref{5D:sugra:STEP2LI:BPS}), and explicitly taking the charge derivatives, we get
\begin{subequations}
\begin{align}
	L&=\fft{\pi^3}{64G_5}(C_{JKL}X^K_\infty X^L_\infty)(C^{JMN}Q_MQ_N)~,\\
	L^I&=\fft{\pi\sqrt3}{2\sqrt{2Z}}(C_{JKL}X^K_\infty X^L_\infty)\left(C^{IJM}Q_M-\fft9{2Z^3}(C^{IMN}Q_MQ_N)(C^{JRS}Q_RQ_S)\right)~,
\end{align}
\end{subequations}
where $Z$ is given in (\ref{eq:Zcc}).  These length scales are the final results of the strong nAttractor mechanism for near-BPS black holes. They reduce to (\ref{5D:STU:L}, \ref{5D:STU:LI}) for the STU model when $C_{123}=1$ and $Q_I>0$.

\subsection{General Extremal Black Holes}
\label{5D:N=2:STEP2:nBPS}

As we have seen, the strong nAttractor mechanism has a straightforward generalization to the case of near-BPS black holes in 5D $\mathcal N=2$ ungauged supergravity.  However, in order to address whether it holds for all near-extremal black holes, we must inquire whether any extremal non-BPS branches exist.  For the STU model in 5D, it can be shown that all extremal solutions are either BPS or related to the BPS solution by flipping the signs of some of the charges.  As the `sign-flipped' solutions retain the linear function structure of $H_I$, it is easily seen that the strong nAttractor mechanism applies to those as well.

In the following, we demonstrate that extremal non-BPS black holes that are disconnected from the extremal BPS branch exist by analyzing the $ST(N)$ family of models. These models have $N$ vector multiplets and are specified by the prepotential
\begin{equation}
	\mathcal V=\fft12X^{N+1}\eta_{ab}X^aX^b\qquad(a,b=1,2,\ldots, N)~,\label{example}
\end{equation}
with $\eta_{ab}=\mathrm{diag}(1,-1,\ldots,-1)$.  The $N$ real scalars parametrize the scalar manifold 
\begin{equation}
	SO(1,1)\times\fft{SO(1,N-1)}{SO(N-1)}~,
	\label{eq:coset}
\end{equation}
that is realized as the hypersurface $\mathcal V=1$ in $(N+1)$-dimensional Riemannian space with coordinates $X^I$. It is known as the 
Jordan symmetric sequence \cite{Gunaydin:1983bi}. For several values of $N$ these models can be obtained by compactification of heterotic 
string theory on $K3\times S^1$  \cite{Cadavid:1995bk,Antoniadis:1995vz}. 

The general extremal black hole solutions to the $ST(N)$ model have not yet been classified, so we first address this before turning to the strong nAttractor mechanism.  Since we are giving up the BPS condition, we must solve the general extremization condition (\ref{5D:sugra:att:eqn:sym}).  Before doing so explicitly, it is instructive to set expectations and consider the group theory of the problem \cite{Cerchiai:2010xv,Ferrara:1997uz,Bellucci:2006xz,Ferrara:2007tu}.

The group $SO(1,1)\times SO(1,N-1)$, identified as the numerator of the coset in (\ref{eq:coset}), is a symmetry of the supergravity theory. It is realized as 
a symmetry of its equation of motion and it preserves the attractor equations (\ref{5D:sugra:att:eqn:sym}). Therefore, given a vector of conserved charges $\vec{Q}=(Q_1,Q_2,\ldots,Q_{N+1})=(Q_a,Q_{N+1})$ that solve the extremization conditions,  we can generate a large orbit of other charge vectors that are also solutions. 
The orbit is 
\begin{equation}
	\mathcal O(\vec{Q})=\fft{SO(1,1)\times SO(1,N-1)}{\mathcal S(\vec{Q})}~,
	\label{eq:orbitQ}
\end{equation}
where $\mathcal S(\vec{Q})$ is the stabilizer of the charge vector $\vec{Q}$ under the action of the group in the numerator.  
In the $ST(N)$ model there are two distinct classes of orbits, depending on the sign of $\eta^{ab}Q_aQ_b$.  In analogy with the Lorentzian interval, we refer to these cases as ``time-like'' when $\eta^{ab}Q_aQ_b$ is positive and ``space-like'' when it is negative.  (We do not consider the ``null'' case because it gives rise to a singular black hole horizon.)  The orbits corresponding to the two classes are given by (\ref{eq:orbitQ}) with respective stabilizer groups \cite{Ferrara:1997uz}
\begin{equation}
\mathcal S(\vec Q)=\begin{cases}
	{SO(N-1)}~, & \vec{Q}\mbox{ is `time-like'}~(\eta^{ab}Q_aQ_b>0)~;\\
	{SO(1,N-2)}~, & \vec{Q}\mbox{ is `space-like'}~(\eta^{ab}Q_aQ_b<0)~.
	\end{cases}
\end{equation}

In general, the stabilizer group acts on the scalars while leaving the charges $\vec Q$ invariant.  However, this will not always generate a new solution, as the scalars live on the coset (\ref{eq:coset}). Rather, the moduli space $\mathcal M(\vec{Q})$ of solutions with a fixed charge vector is given by the quotient space \cite{Cerchiai:2010xv,Bellucci:2006xz,Ferrara:2007tu}
\begin{equation}
	\mathcal M(\vec{Q})=\fft{\mathcal S(\vec{Q})}{\mathrm{m.c.s.}(\mathcal S(\vec{Q}))}~,
\end{equation}
where m.c.s.\ denotes the maximal compact subgroup. We conclude that black holes with time-like and space-like charge vectors feature
the moduli spaces
\begin{equation}
	\mathcal M(\vec{Q})=\begin{cases}
	I~, & \vec{Q}\mbox{ is `time-like'}~(\eta^{ab}Q_aQ_b>0)~;\\
	\displaystyle\fft{SO(1,N-2)}{SO(N-2)}~, & \vec{Q}\mbox{ is `space-like'}~(\eta^{ab}Q_aQ_b<0)~,
	\end{cases}\label{Moduli}
\end{equation}
where $I$ denotes the trivial group. Thus the attractor mechanism for extremal black holes with a time-like charge vector fixes all scalars, while for those with a space-like charge vector it fixes only two scalars and leaves a $(N-2)$-dimensional moduli space on the horizon.

In summary of our group theoretical analysis, we expect the attractor equations (\ref{5D:sugra:att:eqn:sym}) to yield two distinct branches of solutions, depending on whether the charge vector is time-like or space-like.  As shown in Appendix~\ref{App:B}, this is indeed what happens. There we show that, for a given charge vector, there is a unique solution to the attractor equations with $Q_a X^a\neq 0$. In contract, for $Q_a X^a= 0$ the $N$ scalars are merely constrained by two equations, leaving $N-2$ free moduli. As we show below, the first option describes the time-like case while the second possibility is relevant to the space-like case.

\subsubsection{Time-like Charge Vectors and the BPS branch}\label{timelike}

The branch of solutions to the attractor equations (\ref{5D:sugra:att:eqn:sym}) with $Q_aX^a\neq0$ is determined in Appendix~\ref{App:B} as
\begin{subequations}
	\begin{align}
	X^a&=\pm\eta^{ab} Q_bQ_{N+1}\left(\fft12Q_{N+1}\eta^{ab}Q_aQ_b\right)^{-\fft23}~,\label{eq:Xahor}\\
	X^{N+1}&=Q_{N+1}^{-1}\left(\fft12Q_{N+1}\eta^{ab}Q_aQ_b\right)^\fft13~,	\end{align}\label{5D:sugra:ex:att}%
\end{subequations}
along with
\begin{equation}
	R^2=\pm\fft12\left(\fft12Q_{N+1}\eta^{ab}Q_aQ_b\right)^\fft13~.\label{5D:sugra:ex:att:R}
\end{equation}
Since all quantities are real, the cube-roots are unambiguous.  The signs in (\ref{eq:Xahor}) and (\ref{5D:sugra:ex:att:R}) are {\it not} correlated.

These solutions to the attractor equation are not necessarily physical. We must additionally demand that the kinetic function $G_{IJ}$ (\ref{5D:sugra:metric}) for scalar fields $X^I$  and vector fields $A_\mu^I$ in the action (\ref{5D:sugra:action}) must be 
positive definite.  Reading off $C_{IJK}$ from (\ref{example}) and substituting them into the expression for $G_{IJ}$ given in (\ref{5D:sugra:metric}), we obtain the condition
\footnote{We may choose $\phi^a=X^a~(a=1,\cdots,N)$ as $N$ physical scalar fields and compute the corresponding kinetic function $\mathcal G_{ab}$ as $\mathcal G_{ab}=G_{IJ}\partial_{\phi^a}X^I\partial_{\phi_b}X^J$ with $X^{N+1}$ determined in terms of $\phi^a$ by the constraint $\mathcal V=1$. Imposing the positive definite condition on the kinetic function $\mathcal G_{ab}$ then yields the same result.}
\begin{equation}
	\mathrm{Positive~definite}~G_{IJ}\quad\Rightarrow\quad \eta_{ab}X^aX^b>0~,\label{positive:def}
\end{equation}
in this model. For the solution with $Q_aX^a\neq0$ (\ref{eq:Xahor}) gives
\begin{equation}
	\eta_{ab}X^aX^b=2Q_{N+1}\left(\fft12Q_{N+1}\eta^{ab}Q_aQ_b\right)^{-\fft13}~,\label{X^2:timelike}
\end{equation}
which is positive if and only if $\eta^{ab}Q_aQ_b>0$.  This confirms the identification of the $Q_aX^a\neq0$ attractor solutions with the
time-like charge vectors $\eta^{ab}Q_aQ_b>0$ that we indicated in the title of this subsection.

Our explicit computation showing that the attractor equations fix all scalars at the horizon agrees with the expectations from the group theory analysis in the beginning of this subsection.  Moreover, it is instructive to compare this attractor solution to the BPS attractor solution (\ref{eq:BPSattractor}).  To do so, we first compute the constant $Z$ from (\ref{eq:Zcc})
\begin{equation}
Z=3\left(\fft12Q_{N+1}\eta^{ab}Q_aQ_b\right)^{\fft13}~.
\end{equation}
This allows us to rewrite (\ref{5D:sugra:ex:att},\,\ref{5D:sugra:ex:att:R}) as
\begin{equation}
X_a=\pm\fft{Q_a}Z~,\qquad
X_{N+1}=\fft{Q_{N+1}}Z~,\qquad
R^2=\pm\fft{Z}6~.
\label{eq:XXR}
\end{equation}
Comparison with (\ref{eq:BPSattractor}) identifies this as a BPS attractor, provided we choose the positive sign in   (\ref{eq:Xahor}).  In this case, the additional sign choice in (\ref{5D:sugra:ex:att:R}) corresponds to the preservation of one or the other half of the $\mathcal N=2$ supersymmetries.  On the other hand, choosing the negative sign in (\ref{eq:Xahor}) gives rise to an extremal non-BPS solution which is closely related to the BPS solution.

To see the close relation between the solutions with different choice of signs, we note that for the $ST(N)$ model the effective potential $V_{\mathrm{eff}}$ is invariant under the discrete transformations
\begin{subequations}
	\begin{align}
	A:& \quad Q_a\to-Q_a~,\\
	B:& \quad Q_I\to-Q_I~.
	\end{align}\label{sign:flips}%
\end{subequations}
As a result, the attractor equations (\ref{5D:sugra:att:eqn:sym}) are invariant under these same transformations, and hence we can map attractor solutions to one another using them.  The transformation $A$ flips the sign of $X_a$ in (\ref{eq:XXR}), and maps between BPS and non-BPS solutions,  while $B$ flips the sign of $R^2$ in (\ref{eq:XXR}) and maps between BPS solutions preserving opposite sets of supersymmetries.

Since the coupled equations of motion (\ref{5D:sugra:eom}) with $\rho_0=0$ are invariant under these discrete transformations, the full extremal black hole solutions can also be obtained from the BPS black hole solutions (\ref{5D:sugra:BPS},\,\ref{5D:sugra:radial:BPS}) by a combination of $A$ and $B$ transformations.  The solution thus takes the form
\begin{subequations}
	\begin{align}
	X^a&=\pm\eta^{ab} H_bH_{N+1}\left(\fft12H_{N+1}\eta^{ab}H_aH_b\right)^{-\fft23}~,\label{eq:Xa}\\
	X^{N+1}&=H_{N+1}^{-1}\left(\fft12H_{N+1}\eta^{ab}H_aH_b\right)^\fft13~,
	\end{align}\label{5D:BH:att}%
\end{subequations}
along with
\begin{equation}
	R^2=\pm\fft12\left(\fft12H_{N+1}\eta^{ab}H_aH_b\right)^\fft13~.\label{5D:BH:att:r}
\end{equation}
This can be obtained from (\ref{5D:sugra:ex:att},\,\ref{5D:sugra:ex:att:R}) by the replacement $Q_I\to H_I$ where $H_I$ are a set of linear functions defined in (\ref{eq:linfun}), just as in the BPS case. The strong nAttractor mechanism therefore holds for the extremal black hole solutions in the same way as we have seen in section~\ref{5D:N=2:STEP2:BPS}.

To obtain a physical solution, the linear functions should not vanish outside the horizon.  This fixes the signs of the $q_I^\infty$ to match those of the corresponding charges $Q_I$ which in turn ensures that $\eta^{ab}H_aH_b>0$.  The sign in (\ref{5D:sugra:ex:att:R}) and (\ref{5D:BH:att:r}) is then chosen to match the sign of $Q_{N+1}$ so that $R^2$ is positive.

\subsubsection{Space-like Charge Vectors and the Non-BPS branch}
We now turn to the branch of solutions to (\ref{5D:sugra:att:eqn:sym}) with $Q_aX^a=0$.  As we show in Appendix~\ref{App:B}, in this case
the attractor solutions are
\begin{subequations}
	\begin{align}
	\eta_{ab}X^aX^b&=-2Q_{N+1}\left(\fft12Q_{N+1}\eta^{ab}Q_aQ_b\right)^{-\fft13}~,\label{must:spacelike}\\ X^{N+1}&=-Q_{N+1}^{-1}\left(\fft12Q_{N+1}\eta^{ab}Q_aQ_b\right)^\fft13~,
	\end{align}\label{5D:sugra:ex:moduli}%
\end{subequations}
along with
\begin{equation}
	R^2=\pm\fft12\left(\fft12Q_{N+1}\eta^{ab}Q_aQ_b\right)^\fft13~.\label{5D:sugra:ex:moduli:R}
\end{equation}
These attractor solutions differ dramatically from the ones given in (\ref{eq:Xahor}, \ref{5D:sugra:ex:att}, \ref{5D:sugra:ex:att:R}) for the time-like case because here a given charge vector $Q_I$ does not determine $X^I$ unambiguously. It only requires the $N+1$ scalars $X^I$ to satisfy $Q_aX^a=0$ 
and (\ref{must:spacelike}, \ref{5D:sugra:ex:moduli}). The $\mathcal V=1$ constraint is not a separate condition because it is enforced by the last equations taken together.
As a result, only two of the $N$ unconstrained scalars are fixed, leaving a $(N-2)$-dimensional moduli space at the horizon.

The physical stability condition on moduli $\eta_{ab}X^aX^b>0$ given in (\ref{positive:def}) applies independently of the solution we consider. However, 
the relation of this quantity to the charges $Q_I$ that we give in \eqref{X^2:timelike} when assuming $Q_aX^a\neq 0$ differs in sign from our 
result \eqref{5D:sugra:ex:moduli} when $Q_aX^a=0$. Having already argued that the former requires time-like charge vectors $\eta^{ab}Q_aQ_b>0$ we
conclude that the latter, the case considered presently, applies to space-like charge vectors $\eta^{ab}Q_aQ_b<0$.

For space-like charge vectors, a full extremal black hole solution has not yet been constructed. Furthermore, because of the moduli space at the horizon, extending the attractor solution (\ref{5D:sugra:ex:moduli},\,\ref{5D:sugra:ex:moduli:R}) by the simple replacement $Q_I\to H_I$ is ambiguous and therefore we cannot construct a full extremal black hole solution in that way as in \ref{timelike}. Instead, we can implement an appropriate solution generating technique as described in Appendix \ref{App:C}.  The result can still be written in terms of a set of linear functions, (\ref{eq:linfun}), and takes the form
\begin{subequations}
	\begin{align}
	X^a&=\pm H_{N+1}\left(\fft12H_{N+1}\eta^{ab}H_aH_b\right)^{-\fft23}\fft{\eta^{ab}\eta^{cd}H_c(q_d^\infty H_b-q_b^\infty H_d)}{\left(-\fft12\eta^{ab}\eta^{cd}(q_c^\infty H_a-q_a^\infty H_c)(q_d^\infty H_b-q_b^\infty H_d)\right)^\fft12}~,\label{eq:modspsc}\\
	X^{N+1}&=-H_{N+1}^{-1}\left(\fft12H_{N+1}\eta^{ab}H_aH_b\right)^\fft13~,\label{5D:BH:moduli:N+1}
	\end{align}\label{5D:BH:moduli}%
\end{subequations}
along with
\begin{equation}
	R^2=\pm\fft12\left(\fft12H_{N+1}\eta^{ab}H_aH_b\right)^\fft13~.\label{5D:BH:moduli:R}
\end{equation}
The moduli space of scalars at the horizon is manifested in (\ref{eq:modspsc}) through the explicit $q_a^\infty$ factors. Note that the signs in (\ref{eq:modspsc}) and (\ref{5D:BH:moduli:R}) are not correlated.

As in \ref{timelike}, we must fix the signs of the $q_I^\infty$ to match those of the corresponding charges $Q_I$ to obtain a physical solution. This in turn ensures that $\eta^{ab}H_aH_b<0$ for space-like charge vectors. Then the sign in (\ref{5D:sugra:ex:moduli:R}) and (\ref{5D:BH:moduli:R}) has to be opposite of the sign of $Q_{N+1}$ so that $R^2$ is positive.

Since these solutions cannot be obtained from the BPS (extremal) black hole solutions through any choice of sign flips, the analysis of the strong nAttractor mechanism in section~\ref{5D:N=2:STEP2:BPS} no longer applies.  However, since the solution again is given in terms of a set of linear functions $H_I$ (and additionally the constants $q_I^\infty$), the replacement of a radial derivative by a gradient with respect to the charges $Q_I$ contracted with a ``normal vector'' $q_I^\infty$ remains valid.  As a result, the first step in demonstrating the strong nAttractor mechanism continues to hold in these black holes. To be specific, the strong nAttractor mechanism recasts the scales $L$ and $L^I$ as
\begin{subequations}
	\begin{align}
	L&\overset{\scriptscriptstyle \mathrm{standard}}{=}\pi S_{\mathrm{ext}}\lim_{\rho\to0}\left(\fft{\partial R(\rho;0,Q_I,X^I_\infty)^3}{\partial\rho}\right)_{Q_I,X^I_\infty}\nn\\&~\overset{\scriptscriptstyle \mathrm{strong}}{=}\pi S_{\mathrm{ext}}\sum_{I=1}^{N+1}q_I^\infty\fft{\partial R(0;0,Q_I,X^I_\infty)^3}{\partial Q_I}~,\label{5D:sugra:STEP2:radial}\\
	L^I&\overset{\scriptscriptstyle \mathrm{standard}}{=}\fft{2G_5S_{\mathrm{ext}}}{\pi}\lim_{\rho\to0}\left(\fft{\partial X^I(\rho;0,Q_J,X^J_\infty)}{\partial\rho}\right)_{Q_J,X^J_\infty}\nn\\&~\overset{\scriptscriptstyle \mathrm{strong}}{=}\fft{2G_5S_{\mathrm{ext}}}{\pi}\sum_{J=1}^{N+1}q_J^\infty\fft{\partial X^I(0;0,Q_J,X^J_\infty)}{\partial Q_J}~.\label{5D:sugra:STEP2:scalar}
	\end{align}\label{5D:sugra:STEP2}
\end{subequations}

To complete the strong nAttractor picture, we must also determine the ``normal'' vector $q_I^\infty$ in terms of the asymptotic scalars $X^I_\infty$. As in section~\ref{5D:N=2:STEP2:BPS}, we impose the asymptotically free condition, $R^2\to\rho$ as $\rho\to\infty$, to fix the residual scaling symmetry.  This now allows us to invert the relation between $X^I_\infty$ and $q_I^\infty$ given by (\ref{5D:BH:moduli}) in the asymptotic region $\rho\to\infty$, with the result
\begin{subequations}
	\begin{align}
	q_a^\infty&=\fft{2X^{N+1}_\infty(\eta_{ab}X^b_\infty Q_cX^c_\infty-Q_a(\eta_{cd}X^c_\infty X^d_\infty))}{((Q_cX^c_\infty)^2-(\eta^{ab}Q_aQ_b)(\eta_{cd}X^c_\infty X^d_\infty))^\fft12}~,\\
	q_{N+1}^\infty&=\fft{2\,\mbox{sign}[Q_{N+1}]}{X^{N+1}_\infty}~.\label{q:X:N+1}
	\end{align}\label{q:X}
\end{subequations}

Finally, substituting (\ref{q:X}) into (\ref{5D:sugra:STEP2}), we get the length scales written explicitly in terms of conserved charges $Q_I$ and asymptotic scalars $X^I_\infty$. Since the final expressions are quite involved and not illuminating, here we write down the length scale $L$ associated to the entropy only:
\begin{align}
	L=\fft{\pi^3}{64G_5}\biggl(&-(\eta^{ab}Q_aQ_b)(\eta_{cd}X^c_\infty X^d_\infty)\nn\\
	&-4|Q_{N+1}|X^{N+1}_\infty\left((Q_aX^a_\infty)^2-(\eta^{ab}Q_aQ_b)(\eta_{cd}X^c_\infty X^d_\infty)\right)^\fft12\biggr)~.
\end{align}
This completes the strong nAttractor mechanism for general extremal black holes of the $ST(N)$ model.

\section{Discussion}\label{discussion}
The {\it standard} nAttractor mechanism is fairly robust, as we have shown in a variety of examples of section~\ref{standard}. We suspect that generalizations to yet other settings will prove relatively straightforward. The identification embodied in the standard nAttractor mechanism is not only satisfying conceptually, it is also valuable for many practical computations. 

In contrast, we found that the {\it strong} nAttractor mechanism depends more delicately on the particulars of the setup. Our study in section \ref{strong} illustrates this aspect in a family of models that exemplify several distinct structures of the moduli space underlying the attractor mechanism for extremal black holes. A more detailed understanding of how and why ``particulars'' are important would be interesting for the general understanding of holography. 

The {\it standard} nAttractor mechanism realizes the general holographic lore relating the radial flow in a black hole spacetime and a renormalization group flow in manner that adds technical precision by focussing on the final approach to the horizon/IR fixed point. From this holographic point of view it is unsurprising that the standard nAttractor mechanism identifies a thermal derivative (adding energy to the black hole) with a radial derivative in a fixed spacetime (moving away from the IR fixed point). 

The {\it strong} nAttractor mechanism is more ambitious as it seeks an interpretation that is {\it intrinsic} to the end point of the flow by further trading the radial derivative for differentiation in charge space. That it is reminiscent of the Callan-Symanzik equation that identifies a scale transformation in effective quantum field theory with a motion in the space of couplings. 

Technically, the strong nAttractor mechanism follows from the standard one when we can establish a connection between the radial derivative (in the extremal background) and a 
derivative in ``charge  space'' that acts only on the fixed point solution (the near horizon geometry). We find that in favorable cases, such as 4D and 5D BPS black holes, this property follows immediately from the linearity of the ``harmonic'' functions $H_I=Q_I+q_I^\infty\rho$  in the black hole solutions which interpolate between the conserved charges $Q_I$ at the horizon and the normal vector $q_I^\infty$ at infinity. For example, the 5D BPS attractor equation (\ref{eq:BPSattractor}) gives the required connection (\ref{eq:qinfty}) between the normal vector in charge space and the asymptotic scalars. This case exhibits an intricate interplay between the UV and IR where the {\it extremal} attractor mechanism for the scalars determines the normal vector $q_I^\infty$ from the asymptotic scalars. 

These examples are satisfying but generally there clearly can not be a one-to-one map between the radial derivative in physical space and differentiation in charge space. If nothing else, because these spaces do not necessarily have the same dimension when no supersymmetry is preserved by the black hole horizon. 
For example, nonBPS black holes in ${\cal N}=2$ supergravity feature scalar fields that are not fixed by the black hole charges. One might hope to somehow decouple the ``extra'' parameters on one side or the other. Indeed, in the context of BPS black holes, additional ${\cal N}=2$ hyper multiplets and additional scalars in theories with ${\cal N}>2$ supersymmetry decouple. However, it is far from clear that similar decoupling always happens to ``unwanted'' parameters. 

We pursue (but far from fully address) this challenge by investigating the nonBPS branch of black holes in theories that do have BPS black holes.  Specifically, although the relation (\ref{q:X}) follows from the flow realized by the extremal black hole solution (\ref{5D:BH:moduli}), it is unclear whether it can be computed (or somehow rendered irrelevant) solely from extremal horizon considerations.  While the attractor equations that apply in this case (\ref{5D:sugra:ex:moduli}) can not by themselves be inverted to obtain $Q_I$ from $X^I$, it is possible that they can be combined with additional universal input to realize a robust strong nAttractor mechanism, even in the presence of a moduli space.

Rotating black holes present an interesting setting for future study. The standard attractor mechanism has not been fully developed in this context but it almost certainly applies, especially given our example in subsec \ref{CY:BH} and also the more elaborate example in \cite{Larsen:2019oll}. Thus it appears that we can identify the motion in physical space (``away'' from the black hole while staying in the extremal geometry) with an increase in temperature also for rotating black holes. 

However, the strong attractor mechanism is more challenging. Since electric/magnetic charge and angular momentum scale differently, the geometric aspect of the strong nAttractor mechanism that identifies ``away from the horizon'' with ``motion in charge space'' must be adapted to incorporate ``charges'' (including angular momentum) that do not all have the same dimension. Analogous generalizations allowing for simultaneous deformations away from the IR fixed point by multiple operators with distinct dimensions are easily accommodated in quantum field theory, again by the Callan-Symanzik eqution. 

However, they are more challenging in the present context. The space of charges (including angular momentum) generally differs from the moduli space of asymptotic values of scalar fields. Therefore, as in our study of the strong nAttractor mechanism for nonBPS black holes, it appears that a map between them is impossible. The fact that supersymmetric black holes in AdS-spaces must be rotating offers great motivation to overcome such obstacles.

\acknowledgments
We thank Upamanyu Moitra and Sandip Trivedi for discussions. This work was supported in part by the U.S.~Department of Energy under grant DE-SC0007859.

\appendix
\section{4D \texorpdfstring{$\mathcal N=2$}{N=2} Ungauged Supergravity}
\label{App:A}
The action of 4D $\mathcal N=2$ ungauged supergravity takes the general form (\ref{4D:action}) but further specifies $G_{i\bar j}$, $\mu_{IJ}$, and $\nu_{IJ}$ in 
terms of a holomorphic prepotential $F(X^I)$ as
\begin{align}
	G_{i\bar j}&=\partial_i\partial_{\bar j}\mathcal K~,\\
	\nu_{IJ}-i\mu_{IJ}&=\bar F_{IJ}+2i\fft{(\Im F_{IK})X^K(\Im F_{JL})X^L}{X^K(\Im F_{KL})X^L}~.
\end{align}
We have introduced $X^I$ that parametrize the scalar fields $z^i$ as $z^i=x^i-iy^i=X^i/X^0$ and the K$\ddot{a}$hler potential $\mathcal K$ as 
\begin{equation}
	\mathcal K=-\log[i(\bar X^IF_I-X^I\bar F_I)].
\end{equation}
Here $F_I$ and $F_{IJ}$ denote $\partial_{X^I}F$ and $\partial_{X^I}\partial_{X^J}F$, respectively. 

In this supergravity theory, the effective potential $V_{\mathrm{eff}}$ introduced in (\ref{4D:eff:potential}) can be written concisely as
\begin{align}
	V_{\mathrm{eff}}&=2\left(|Z|^2+4G^{i\bar j}\partial_i|Z|\partial_{\bar j}|Z|\right)~,\label{N=2:4D:potential}
\end{align}
where we have defined the spacetime central charge $Z$ as
\begin{equation}
	Z=e^{\mathcal K/2}(P^I F_I  - Q_I X^I)~.
\end{equation}
The spacetime central charge $Z$ is also useful in writing down the BPS conditions satisfied by all supersymmetric black holes. 
The BPS conditions for the general spherically symmetric extremal black hole ansatz (\ref{4D:ansatz}) with $r_0=0$, are given as
\begin{subequations}
\begin{align}
	|Z|&=\sqrt2(R-r\partial_rR~),\\
	\partial_i|Z|&=\fft{1}{\sqrt2}rRG_{i\bar j}\partial_r\bar z^{\bar j}~.
\end{align}\label{4D:BPS}%
\end{subequations}
It is straightforward to check that these BPS conditions (\ref{4D:BPS}) satisfy the following equations of motion derived from the action (\ref{4D:action}) for the ansatz (\ref{4D:ansatz}) with $r_0=0$ automatically:
\begin{subequations}
\begin{align}
	0=&~\partial_r((r^2-r_0^2)G_{i\bar j}\partial_r\bar z^{\bar j})-(r^2-r_0^2)(\partial_iG_{j\bar k})\partial_rz^j\partial_r\bar z^{\bar k}-\fft{\partial_iV_{\mathrm{eff}}}{4R^2}~,\label{4D:scalar}\\
	0=&~\partial_r((r^2-r_0^2)\partial_r\log R)-1+\fft{V_{\mathrm{eff}}}{4R^2}~,\label{4D:Einstein:1}\\
	0=&~\fft{\partial_r^2 R}{R}+G_{i\bar j}\partial_r z^i\partial_r\bar z^{\bar j}~.\label{4D:Einstein:2}
\end{align}\label{4D:eom}
\end{subequations}
%

\section{Solution of the Attractor Equations for the \texorpdfstring{$ST(N)$}{ST(N)} model.}
\label{App:B}
The attractor equations (\ref{5D:sugra:att:eqn:sym}) for the prepotential (\ref{example}) with the constraint $\mathcal V=1$ are given explicitly as
\begin{subequations}
	\begin{align}
	0&=2(Q_bX^b)Q_a-\eta_{ab}X^b(\eta^{cd}Q_cQ_d - \lambda X^{N+1})~,\label{ex:att:1}\\
	0&=2X^{N+1}Q_{N+1}^2+\fft12\lambda\eta_{ab}X^aX^b~,\label{ex:att:2}\\
	1&=\fft12X^{N+1}\eta_{ab}X^aX^b~,\label{ex:att:3}
	\end{align}\label{ex:att}%
\end{subequations}
where $a=1,2,\cdots,N$. Here $\eta_{ab}$ and $\eta^{ab}$ are \emph{not} lowering and raising operators: we do not raise or lower indices at all in this Appendix.

The constraint \eqref{ex:att:3} shows that $X^{N+1}\neq 0$ and combines with \eqref{ex:att:2} to give the Lagrange multiplier
\begin{equation}
	\lambda=-2Q_{N+1}^2(X^{N+1})^2~.\label{step:lagmult}
\end{equation}
Contracting \eqref{ex:att:1} with $X^a$ and simplifying using the constraint \eqref{ex:att:3} we further have 
\begin{equation}
	(Q_aX^a)^2 =   \fft{\eta^{ab}Q_aQ_b}{X^{N+1}}-\lambda =  \frac{1}{X^{N+1}} \left( \eta^{ab}Q_aQ_b  + 2Q_{N+1}^2(X^{N+1})^3\right)
	 ~.\label{eqn:QXeqn}
\end{equation}
On the other hand, contracting with $\eta^{ab}Q_b$ gives
\begin{equation}
(Q_cX^c)\left(\eta^{ab}Q_aQ_b+\lambda X^{N+1}\right)=0~.
\label{eq:QXQQ}
\end{equation}
To make further progress we must make an assumption on whether the inner product $Q_aX^a$ vanishes. With foresight, we will refer to 
the case $Q_a X^a\neq 0$ as timelike and the case $Q_a X^a = 0$ as spacelike.

\subsection{Time-like Solutions: \texorpdfstring{$Q_a X^a\neq 0$}{Q.X ne 0}}

If we assume that $Q_a X^a\neq 0$, then (\ref{eq:QXQQ}) leads directly to
\begin{equation}
	\lambda=-\fft{\eta^{ab}Q_aQ_b}{X^{N+1}}~.\label{eqn:lambda}
\end{equation}
Combining this with \eqref{step:lagmult} then gives the scalar $X^{N+1}$ in terms of the charges
\begin{align}
	X^{N+1}=\left( \fft{\eta^{ab}Q_aQ_b}{2Q_{N+1}^2}\right)^{\fft13}  ~.\label{eqn:XN1time}
\end{align}
Now \eqref{eqn:QXeqn} gives
\begin{equation}
	(Q_aX^a)^2  = \fft{2\eta^{ab}Q_aQ_b}{X^{N+1}}    = \left( 4\eta^{ab}Q_aQ_b ~Q_{N+1} \right)^{\fft23}   ~,\label{eqn:QXeqn2}
\end{equation}
and finally \eqref{ex:att:1} simplified using \eqref{eqn:lambda} gives the attractor values of the remaining scalars
\begin{equation}
X^a  =\eta^{ab} Q_b   \fft{Q_c X^c}{\eta^{de}Q_dQ_e}  = \pm \eta^{ab} Q_b  \fft{(4Q_{N+1})^\fft13}{(\eta^{cd}Q_cQ_d)^\fft23} ~.
\end{equation}

\subsection{Space-like Solutions: \texorpdfstring{$Q_aX^a=0$}{Q.X = 0}}

If on the other hand $Q_a X^a = 0$, the first equation in \eqref{eqn:QXeqn} gives the Lagrange multiplier  
\begin{equation}
	\lambda=\fft{\eta^{ab}Q_aQ_b}{X^{N+1}}~,\label{eqn:lambda2}
\end{equation}
while the second one gives the attractor value for the scalar
\begin{align}
	X^{N+1}= - \left( \fft{\eta^{bc}Q_bQ_c}{2Q_{N+1}^2}\right)^\fft13  ~.\label{eqn:XN1space}
\end{align}
Importantly, in this case \eqref{ex:att:1} requires only \eqref{eqn:lambda2} and imposes nothing further on the scalars $X^a$. 
Therefore, for space-like solutions  the scalars are only constrained by the two conditions $Q_aX^a=0$ and $\mathcal V=1$. This means there is a $(N-2)$-dimensional moduli space at the horizon. For $N=2$ the $ST(N)$ model reduces to the $STU$ model which indeed has no moduli space.

\section{Deriving Black Hole Solutions for ``Space-like'' Charge Vectors}
\label{App:C}

In this appendix, we derive an extremal black hole solution to (\ref{5D:sugra:eom}) with $\rho_0=0$ for a `space-like' charge vector $\eta^{ab}Q_aQ_b<0$ in the $ST(N)$ model. For simplicity, we choose $N=3$ here and therefore $a,b=1,2,3$ and $I,J=1,2,3,4$. The case with general $N$ can be studied in a similar way.

For $N=3$, we make use of the symmetry $SO(1,1)\times SO(1,2)$ corresponding to the top of the coset (\ref{eq:coset}) to generate a more general solution from a known seed solution.  To start with, we take $X^3=Q_3=0$, in which case the effective potential (\ref{eq:Veff}) and the cubic prepotential (\ref{eqn:prepot}) for the $ST(N)$ model reduce to
\begin{align}
	V_{\rm eff}&=(X^+)^2Q_+^2+(X^-)^2Q_-^2+(X^4)^2Q_4^2~,\\
	\mathcal V&=X^+X^-X^4~,
\end{align}
where we have defined $X^\pm=\fft{1}{\sqrt2}(X^1\pm X^2)$ and $Q_\pm=\fft{1}{\sqrt2}(Q_1\pm Q_2)$. Since these potentials are exactly the same as those of STU model, we know that the extremal black hole solutions to (\ref{5D:sugra:eom}) with $\rho_0=0$ must take the form of extremal STU black holes when $X^3=Q_3=0$. For a `space-like' charge vector, they are given explicitly as
\begin{subequations}
	\begin{align}
	R&=\pm 2^{-\fft12}\left(\fft12 H_4\left(H_1^2-H_2^2\right)\right)^\fft16~,\\
	\fft{1}{\sqrt2}(X^1+X^2)=X^+&=\pm\fft{(H_+H_-H_4)^\fft13}{H_+}=\pm\fft{\left(\fft12 H_4\left(H_1^2-H_2^2\right)\right)^\fft13}{(H_1+H_2)/\sqrt2}~,\label{Xp}\\
	\fft{1}{\sqrt2}(X^1-X^2)=X^-&=\mp\fft{(H_+H_-H_4)^\fft13}{H_-}=\mp\fft{\left(\fft12 H_4\left(H_1^2-H_2^2\right)\right)^\fft13}{(H_1-H_2)/\sqrt2}~,\label{Xm}\\
	X^3&=0~,\\
	X^4&=-\fft{(H_+H_-H_4)^\fft13}{H_4}=-\fft{\left(\fft12 H_4\left(H_1^2-H_2^2\right)\right)^\fft13}{H_4}~,
	\end{align}\label{Seed}%
\end{subequations}
where $H_1,H_2,H_4$ are the linear functions given as (\ref{eq:linfun}) and $H_\pm=\fft{1}{\sqrt2}(H_1\pm H_2)$. The correlated signs between (\ref{Xp}) and (\ref{Xm}) are essential for $\eta_{ab}X^aX^b=(X^1)^2-(X^2)^2>0$ to be satisfied under the `space-like' charge condition $\eta^{ab}Q_aQ_b=Q_1^2-Q_2^2<0$. See (\ref{positive:def}) to understand why $\eta_{ab}X^aX^b$ has to be positive.

We want to generate extremal black hole solutions without the constraints $X^3=Q_3=0$, starting from the seed solution (\ref{Seed}). Hence we need an appropriate transformation that yields a new solution from a given one.  Since the coupled equations of motion (\ref{5D:sugra:eom}) are invariant under $SO(1,1)\times SO(1,2)$, we use the latter symmetry to map $X^I,R,Q_I$ to the transformed $\tilde X^I,\tilde R,\tilde Q_I$ according to
\begin{equation}
	\tilde X^a=(\Lambda^{-1})^a{}_b X^b~,\quad \tilde X^4=X^4~,\quad \tilde R=R~,\quad \tilde Q_a=\Lambda^b{}_a Q_b~,\quad \tilde Q_4=Q_4~,\label{Generating}
\end{equation}
where $\Lambda\in SO(1,2)$.  Using this symmetry, we can generate an extremal black hole solution for general `space-like' charge vectors with one-dimensional moduli space at the horizon, starting from the seed solution (\ref{Seed}). 

To be more explicit, we must determine the transformation matrix $\Lambda$ that yields such a black hole solution. Here we split the transformation matrix $\Lambda$ into two parts as $\Lambda(\phi,\theta)=\Lambda_1(\phi)\Lambda_2(\theta)$ with two free parameters $\phi$ and $\theta$: $\Lambda_1(\phi)$ denotes a $SO(1,2)$ transformation preserving the original charge vector $(Q_1,Q_2,0)$, which generates a one-dimensional moduli space; $\Lambda_2(\theta)$ denotes a $SO(2)$ transformation on the 23-plane, which generalizes the charge vector to $(\tilde Q_1,\tilde Q_2,\tilde Q_3)$. These matrices can be written explicitly as 
\begin{subequations}
\begin{align}
	\Lambda(\phi,\theta)&=\Lambda_1(\phi)\Lambda_2(\theta)~,\\
	\Lambda_1(\phi)&=
	\begin{pmatrix}
	\cosh\gamma & \sinh\gamma & 0 \\
	\sinh\gamma & \cosh\gamma & 0 \\
	0 & 0 & 1
	\end{pmatrix}^{-1}
	\begin{pmatrix}
	\cosh\phi & 0 & \sinh\phi \\
	0 & 1 & 0 \\
	\sinh\phi & 0 & \cosh\phi
	\end{pmatrix}
	\begin{pmatrix}
	\cosh\gamma & \sinh\gamma & 0 \\
	\sinh\gamma & \cosh\gamma & 0 \\
	0 & 0 & 1
	\end{pmatrix}~,\\
	\Lambda_2(\theta)&=
	\begin{pmatrix}
	1 & 0 & 0 \\
	0 & \cos\theta & -\sin\theta \\
	0 & \sin\theta & \cos\theta
	\end{pmatrix}~,
\end{align}\label{Generating:matrix}%
\end{subequations}
where we have defined $\gamma$ as $\sinh\gamma=\mbox{sign}[Q_2]Q_1/\sqrt{Q_2^2-Q_1^2}$. Note that the boost matrices with the fixed parameter $\gamma$ play a role of transforming the charge vector between $(Q_1,Q_2,0)$ and $(0,\mbox{sign}[Q_2]\sqrt{Q_2^2-Q_1^2},0)$ and the one with a free parameter $\phi$ corresponds to the boost preserving the transformed charge vector $(0,\mbox{sign}[Q_2]\sqrt{Q_2^2-Q_1^2},0)$. In short, we have $Q_a=\Lambda_1(\phi)^b{}_aQ_b$.

Transforming the seed solution (\ref{Seed}) as (\ref{Generating}) with the transformation matrix $\Lambda(\phi,\theta)$ given in (\ref{Generating:matrix}) then yields an extremal black hole solution specified by $\tilde R,\tilde X^I$ for general `space-like' charge vectors $(\tilde Q_1,\tilde Q_2,\tilde Q_3)=(Q_1,Q_2\cos\theta,-Q_2\sin\theta)$. In particular, the resulting black hole solution will have a one-dimensional moduli space at the horizon parametrized by a free parameter $\phi$. 

The resulting black hole solution, however, is written in terms of four free parameters $q_1^\infty,q_2^\infty,q_4^\infty,\phi$ and the corresponding expression is quite involved and not illuminating. With foresight that the final expression can be simplified in terms of the linear functions $\tilde H_I=\tilde q_I^\infty\rho+\tilde Q_I$ with the transformed charges $\tilde Q_I$, we introduce alternative free parameters $\tilde q_I^\infty$ as $\tilde q_a^\infty=\Lambda^b{}_aq_b^\infty$ and $\tilde q_4^\infty=q_4^\infty$ following the charge transformation in (\ref{Generating}), where we set $q_3^\infty$ as 0. Then we have the following two constraints,
\begin{equation}
	0=Q_3=(\Lambda^{-1})^b{}_3\tilde Q_b~,\qquad 0=q_3^\infty=(\Lambda^{-1})^b{}_3\tilde q_b^\infty~,\label{Constraint}
\end{equation}
which determine $\phi$ as
\begin{align}
	\sinh\phi=\fft{\mathrm{sign}[\tilde Q_2(\tilde q_1^\infty\tilde Q_2-\tilde q_2^\infty\tilde Q_1)+\tilde Q_3(\tilde q_1^\infty\tilde Q_3-\tilde q_3^\infty\tilde Q_1)](\tilde q_3^\infty\tilde Q_2-\tilde q_2^\infty\tilde Q_3)\sqrt{\tilde Q_2^2+\tilde Q_3^2-\tilde Q_1^2}}{\mbox{sign}[Q_2]\sqrt{(\tilde Q_2^2+\tilde Q_3^2)((\tilde q_1^\infty\tilde Q_2-\tilde q_2^\infty\tilde Q_1)^2+(\tilde q_1^\infty\tilde Q_3-\tilde q_3^\infty\tilde Q_1)^2-(\tilde q_2^\infty\tilde Q_3-\tilde q_3^\infty\tilde Q_2)^2)}}.\label{Parameters}
\end{align}
Substituting $q_a^\infty=(\Lambda^{-1})^b{}_a\tilde q_b^\infty$, $q_4^\infty=\tilde q_4^\infty$, and (\ref{Parameters}) into the black hole solution obtained by the solution generating technique described above, we can rewrite it in terms of the free parameters $\tilde q_I^\infty$ instead of $q_1^\infty,q_2^\infty,q_4^\infty,\phi$. The final expression, when generalized to arbitrary $N$, is then given as (\ref{5D:BH:moduli},\,\ref{5D:BH:moduli:R}) where we remove the tilde that we have used to distinguish the transformed solution from the seed solution.

\bibliographystyle{JHEP}
\bibliography{nAttractor}

\end{document}